\documentclass[conference]{IEEEtran}
\IEEEoverridecommandlockouts
\usepackage{cite}
\usepackage{amsmath,amssymb,amsfonts}
\usepackage{algorithmic}
\usepackage[]{subfigure}
\usepackage{multirow}
\usepackage{graphicx}
\usepackage{textcomp}     
\usepackage{makecell}
\usepackage{eurosym}
\usepackage{tikz}
\usetikzlibrary{arrows,shapes,positioning,shadows,trees}
\usetikzlibrary{shapes.geometric, arrows}        
\usepackage{forest}
\usetikzlibrary{mindmap,trees}             
\usepackage{xcolor}
\usepackage[ruled,vlined]{algorithm2e}
\usepackage{float}
\usepackage{array}
\usepackage{multirow,multicol}
\usepackage{eurosym}
\usepackage{siunitx}
\usepackage{booktabs}
\usepackage{caption}
\usepackage{tikz}
\newbox\tempbox%
\def\BibTeX{{\rm B\kern-.05em{\sc i\kern-.025em b}\kern-.08em
    T\kern-.1667em\lower.7ex\hbox{E}\kern-.125emX}}
    
\begin{document}

\title{A Comprehensive Multi-Period Optimal Power Flow Framework for Smart LV Networks
\thanks{The authors acknowledge the funding from Luxembourg National Research
Fund  (FNR)  in  the  framework  of  gENESiS  project  (C18/SR/12676686).  I.
Avramidis and F. Capitanescu are with Luxembourg Institute of Science and
Technology (iason.avramidis, florin.capitanescu@list.lu); G. Deconinck is with
KU Leuven Belgium (geert.deconinck@kuleuven.be).}
}

\author{\IEEEauthorblockN{Iason I. Avramidis, \textit{Student Member, IEEE}, Florin Capitanescu, and Geert Deconinck,  \textit{Senior Member, IEEE}}}
\maketitle

\begin{abstract}
This paper presents an extensive multi-period optimal power flow framework, with new modelling elements, for smart LV distribution systems that rely on residential flexibility for combating operational issues. A detailed performance assessment of different setups is performed, including: ZIP flexible loads (FLs), varying degrees of controllability of conventional residential devices, such as electric vehicles (EVs) or photovoltaics (PVs), by the distribution system operator (DSO) (adhering to customer-dependent restrictions) and full exploitation of the capabilities offered by state-of-the-art inverter technologies. A comprehensive model-dependent impact assessment is performed, including phase imbalances, neutral and ground wires and load dependencies. The de-congestion potential of common residential devices is highlighted, analyzing capabilities such as active power redistribution, reactive power support and phase balancing. Said potential is explored on setups where the DSO can make only partial adjustments on customer profiles, rather than (as is common) deciding on the full profiles. The extensive analysis can be used by DSOs and researchers alike to make informed decisions on the required levels of modelling detail, the connected devices and the degrees of controlability. The formulation is computationally efficient, scaling well to medium-size systems, and can serve as an excellent basis for building more tractable or more targeted approaches.
\end{abstract}

\begin{IEEEkeywords}
Multi-Period Optimal Power Flow, Residential Flexibility, Smart Distribution Systems, Unbalanced  Systems
\end{IEEEkeywords}

\section*{Nomenclature}

\subsection{Sets}
\begin{IEEEdescription}%[\IEEEusemathlabelsep\IEEEsetlabelwidth{$V_1,V_2,V_3$}]
\item[${\cal E}$] Set of electric vehicles (EVs)
\item[${\cal F}, {\cal Z}$] Sets of ``phases": ${\cal Z}$ = \{a,b,c\},  ${\cal F}$ = ${\cal Z} \cup$\{n,g\} 
\item[${\cal I}$] Set of nodes
\item[${\cal L}$] Set of flexible loads (FLs) 
\item[${\cal P}$] Set of photovoltaics (PVs) 
\item[${\cal T}$] Set of time periods
\item[${\cal T}^{nc}_{e}$] Set of time periods when EV $e$ may not charge 
\end{IEEEdescription}

\subsection{Parameters}
\begin{IEEEdescription}
\item[$c^{FL}$] Price of FL active power modification, \euro/kWh
\item[$c^{DS}$] Price of EV active discharge to grid, \euro/kWh
\item[$c^{EV}$] Price of EV active charge alteration, \euro/kWh
\item[$c^{I/E}$] Price of import/export from/to MV level, \euro/kWh
\item[$c^{PPV}$] Price of PV active production curtailment, \euro/kWh
\item[$c^{QPV}$] Price of PV reactive capability utilization, \euro/kVar
\item[$c^{v}$] Penalty for technical limits violation, \euro/p.u.
%\item[$\eta^{EV}$] \textcolor{blue}{EV battery conversion efficiency, \%}
\item[$M^{FL}$] Maximum FL alteration, \%
\item[$M^{PV}$] Maximum PV curtailment, \%
\item[$S^{gen}_{p,z,t}$] PV apparent power generated, phase $z$, period $t$, p.u.
\item[$P^{0}_{e,t}$] EV original active charge, phase $z$, period $t$, p.u.
\item[$R_{ij, f\theta}$] Resistance of branch $ij$, between phases $f$, $\theta$, p.u.
\item[$X_{ij, f\theta}$] Reactance of branch $ij$, between phases $f$, $\theta$, p.u. 
\end{IEEEdescription}

\subsection{Variables}
\begin{IEEEdescription}%[\IEEEusemathlabelsep\IEEEsetlabelwidth{$V_1,V_2,V_3$}]
\item[$P^{I/E}_{z,t}$] Active power import/export, phase $z$, period $t$, p.u.
\item[$S^{inv}_{p,z,t}$] PV inverter apparent power, phase $z$, period $t$, p.u.
\item[$P^{inj}_{p,z,t}$] PV active grid injection, phase $z$, period $t$, p.u.
\item[$Q^{PV}_{p,z,t}$] PV reactive power, phase $z$, period $t$, p.u.
\item[$P^{Oc}_{e,z,t}$] EV active ``overcharge", phase $z$, period $t$, p.u.
\item[$P^{Uc}_{e,z,t}$] EV active ``undercharge", phase $z$, period $t$, p.u.
\item[$P^{ds}_{e,z,t}$] EV active discharge, phase $z$, period $t$, p.u.
%\item[$i^{D,P}_{i,f,t}$] Active load current,bus $i$, phase $f$, period $t$, p.u.
%\item[$i^{D,Q}_{i,f,t}$] Reactive load current, bus $i$, phase $f$, period $t$, p.u. 
\item[$P^{D}_{i,z,t}$] Active load demand, bus $i$, phase $z$, period $t$, p.u. 
\item[$Q^{D}_{i,z,t}$] Reactive load demand, bus $i$, phase $z$, period $t$, p.u. 
\item[$Q^{I/E}_{z,t}$] Reactive power import/export, phase $z$, period $t$, p.u.
\item[$u_{i,f,t}$] Voltage magnitude, bus $i$, phase $f$, period $t$, p.u. 
\item[$\sigma_{i,z,t}^{up}$] Overvoltage violation, bus $i$, phase $z$, period $t$, p.u. 
\item[$\sigma_{i,z,t}^{down}$] Undervoltage violation, bus $i$, phase $z$, period $t$, p.u.
\item[$\sigma_{ij,z,t}$] Thermal violation, branch $ij$, phase $z$, period $t$, p.u.
\end{IEEEdescription}

\section{Introduction}

\subsection{Motivation}

In adhering with the smart grid vision, distribution systems are transforming into evermore active systems, characterized by high shares of distributed energy resources (DERs), and high degrees of operational controllability by distribution system operators (DSOs) \cite{SGVision}. The proper management of such distribution systems is crucial; having been designed under the (now) archaic philosophy of fit-and-forget, they are usually ill-equipped to handle the uncoordinated, large-scale integration of DERs \cite{Passive_to_Active}. Especially in LV networks, which are inherently unbalanced, operational issues are usually more prevalent and of elevated severity. For the highest penetration levels, the stress inflicted on such systems can result in harmful voltage spikes or dips and damaging thermal loading of the distribution equipment, all of which are difficult to effectively contain  \cite{DSF_AVRA}. 

To fully understand the benefits of residential flexibility resources (FRs), detailed models of the various devices and the distribution systems themselves are needed. Points of interest include the impact of the load modelling detail on the operational profile, the behavior of the neutral and ground voltages (for protection studies) and the interactions between differently loaded phases. However, most research works utilize simplified load and network models, or/and convex relaxations, targeting scalability rather than accuracy. 

\subsection{Literature review}

Works that opt for solving exact (i.e., non-relaxed) formulations usually make non-generic, case-specific simplifications, such as ignoring the neutral wire or assuming small load imbalances to name a few. The seminal paper on multi-period optimal power flow (MP-OPF) for active distribution systems, \cite{DOPF}, employed the single-phase (1$\Phi$) network representation and the constant P/Q load model. While subsequent papers have since introduced more advanced models for both MV and LV systems, see \cite{4-wire_model}, most papers prioritize the solution technique rather than the model, assuming that non-generic, case-specific simplifications are always expected to hold.

The MP-OPF problems between the MV and LV level share some conceptual similarities, such as the radial network structure, the unbalanced system conditions or the non-negligible impact of line resistance/reactance. In MV systems, the DSO has various resources at its disposal, such as capacitors banks, network switches (reconfiguration), distributed generators and tap changers, see \cite{Efficient_OLTC, Reconfiguration_FC, OPF_SmartDistributionFeeders}. However, the situation is very different in LV systems, where the DSO has far less controllability and equipment available. System management is achieved primarily though electric vehicles (EVs) and photovoltaics (PVs), and rarely through energy storage (ES) systems.

In terms of MP-OPF features, the authors of \cite{Karagiannopoulos_PSCC} eliminate the neutral phase through Kron reduction and calculate an optimal control strategy through an iterative approach (constant P/Q load).  The technique is also used in \cite{CVR_Unbalanced}, where the current-based formulation is employed instead (simultaneous study of MV and LV network). The work \cite{MPOPF_CurrentImbalance} proposes a current-mismatch MP-OPF to optimize a feeder's operation, ignoring the grounding and assuming full controlability of residential ES systems by the DSO (constant P/Q load).

In \cite{Local_improvement_EV_PV}, a local EV charging strategy is applied with respect to the PV's operation, though assuming that the EV is available at all times (constant P/Q load, 3-phase network). A multi-period approach is proposed in \cite{Rolling_MPOPF} for designing the entire charging strategy of some EVs, subject to to dynamic electricity pricing (current mismatch formulation, 3-phase network). The authors of \cite{Central_and_Local_DS} combine central and local control strategies for managing distribution systems through PV utilization (constant P/Q load, neutral consideration).

Approaches based on PVs and EVs that are more specialized have also been proposed. For example, in \cite{PV_4wire}, the 3-stage, centralized, single-period PV curtailment and reactive management problem is solved for 24 consecutive hours for a four-wire LV distribution system (constant P/Q load). The authors of \cite{PhaseBalancing} employ three-phase (3$\Phi$) inverters for phase balancing in pure 3$\Phi$ networks (constant P/Q load). The usage of ES is more rarely tackled, due to their low penetrations in LV networks and the undesirable level DSO involvement in household equipment management. In \cite{OptimalBSSDispatchUnbalanced}, the centralized control of ES in unbalanced networks is studied, though their temporal constraints are ignored (neutral consideration, constant P/Q load). The authors of \cite{BSS_LV} present a highly detailed ES model specifically for LV systems and propose a local area control strategy involving several network agents (Kron reduction, constant P/Q load).

While the constant P/Q load model is the one most commonly employed, more intricate models have occasionally been explored. Explicit ZIP coefficients for commercial, residential and industrial loads have been proposed for conservation of voltage reduction (CVR) studies in unbalanced distribution systems \cite{CVR_Unbalanced_High_PV}, though this customer variety is only found in MV systems. A common ZIP load structure has even been employed in studies following model-free approaches of distribution system optimization, see \cite{Model-Free_Unbalanced} for example (phasor regulation strategy), with the work assuming the simplest form of an unbalanced system. Comparisons between different load models have been performed, though these are either on a purely technical level, i.e., behavior of single, standalone load \cite{ZIP_CVR}, or comparisons between the standard ZIP model and different approximations of it \cite{Impact_of_load_model_unbalanced}. Comprehensive studies of all possible ZIP structures as they related to optimizing distribution system behavior are lacking from the literature.

In terms of solution techniques, convex relaxations are often employed for multi-phase networks. The second-order cone programming (SOCP) and semi-definite programming (SDP) relaxations are popular choices that have spawned different variations (based on types of connection and imbalance), though the neutral wire and ground are rarely included \cite{OPF_Unbalanced_Low, OPF_Unbalanced_DeltaLow}. The branch flow formulation is also commonly employed in radial systems \cite{BranchFlowPartOne}. While extendable to unbalanced systems, it usually includes power flow linearizations or assumptions of very small imbalances \cite{CapacityAssessmentUnbalanced, ReactiveDispatchDistribution}. However, most relaxations hold very rarely for realistic power systems \cite{ACOPF_DS}.

Approximations for specialized versions of the unbalanced MP-OPF have also been proposed. The authors of \cite{UnbalancedWithNeutralSDP} employ the SDP relaxation for distribution systems with neutral cables and fixed ZIP loads. The work \cite{UOPF_MILP_MIQCP} proposes linear and  quadratic simplifications of an MI optimization problem that addresses different operating conditions between MV and LV systems (exponential ZIP load). Aside from the simplifications, these formulations are not entirely practical in representing the behavior of realistic distribution networks. 

In recapitulating, all proposed approaches (simplified or not) do produce promising results. However, if the solved problem is a relaxed one, the original system is largely simplified and the results are rarely feasible/reliable. On the other hand, in exact approaches, the neutral and ground cables are usually ignored, while the effects of the mutual couplings and the load types are not always properly represented.

\subsection{Contributions \& Paper structure}

To focus on scalability, most works sacrifice some accuracy, leaving the behavior of modern (vast flexibility options array) and realistic LV systems in MP settings insufficiently addressed. This work develops a comprehensive, versatile and easily reproducible MP-OPF tool for unbalanced distribution systems and realistic device models. 

This work draws inspiration from and significantly extends several past works. The current-based formulation is a combination of \cite{Karagiannopoulos_PSCC, CVR_Unbalanced, OptimalRestorationUnbalanced}, extended to include the interactions with the neutral and ground wires and the rotation of each phase to a common reference plane. The load modelling is adapted from \cite{4-wire_model}, also accounting for partial load flexibility. The PV modelling is adapted from \cite{4-wire_model, Karagiannopoulos_PSCC}, also including limited curtailment capability and a realistic production capability curve (PCC). The modelling of EV flexibility is original. For the phase balancing through 3$\Phi$ inverters, this work extends \cite{PhaseBalancing}, in accounting for the re-scheduling of the user-driven charging profile. The proposed approach for remunerating customer participation is also a novel addition.

On top of the novel modelling elements, the more conceptual contributions of the paper can be summarized as follows:

\begin{itemize}
    \item It provides an extensive analysis of a comprehensive set of available modelling decisions (many often disregarded) for the optimal management of LV distribution systems.
    \item It construct a generic MP-OPF model that can provide valuable information on how to unlock the full flexibility potential of common LV networks. The framework is easily reproducible, adaptable to each researcher's needs and ideal as a basis for more sophisticated developments.
    \item As the topic is underaddressed, the paper proposes a first crude DSO/customer collaboration framework through which the DSO can utilize residential devices to achieve better system management. The importance of the framework is paramount, given the DSO's traditionally minimal involvement in managing LV systems. 
\end{itemize}

The great advantage of the developed tool is its adaptability to the specific needs of each problem (load/network/device model). It can be used under a vast array of optimization setups, including active power redistribution, reactive power management schemes or phase balancing. This offers great insight to DSOs, which can explore the behavior of often neglected system parts in a reliable manner, and unlock the network's full potential for residential flexibility utilization. Informed decisions can be made on which modelling elements are required and which can be reliably ignored.  

The remainder of this paper is structured as follows. The problem formulation and the main assumptions are presented in Section II. The case study is extensively analyzed in Section III. Conclusions are drawn in Section IV. 

\section{Problem Formulation}

\subsection{Problem assumptions}

For the sake of clarity, we lay out the main problem assumptions. This is day-ahead (planning), multi-period (24-hour horizon, hourly resolution), centralized control optimization problem, where the DSO has partial controllability of the available flexibility resources (FRs). The work is contained to the deterministic setting, assuming the DSO uses a most-likely-to-occur forecast scenario for its planning (real-time deviations are addressed on-the-spot, though this is out of scope). We assume the existence of a digital platform through which customers inform the DSO of their ideal controllable device schedules, initially designed through a rule-based approach or by a sophisticated software (e.g., energy management system). Each customer may or may not receive new set-points for their controllable devices. The expected difference (not the real-time deviations) between the original customer profile and the designed, post-request profile, is the basis for the customer's remuneration. All the necessary software is pre-installed. The generic formulation is applicable for most setups. 

This work covers radial, LV distribution systems. Given the proposed framework's generic structure, it can be used to model 4-wire, multi-grounded systems (found in North America, Europe), 3-wire, grounded or ungrounded systems (found in Europe, UK), single-wire with earth return (found in Australia), and cases of highly specialized grounding (high resistance/reactance, Petersen coils) \cite{EDN_Book}. More intricate configurations, e.g., 5-wire systems, are out of scope.

This work fits within the generic MP-OPF formulation originally developed for 1$\Phi$ networks representations in \cite{DOPF}. The proposed formulation employs the rectangular coordinates formulation for voltages and currents \eqref{VoltageSplit}-\eqref{CurrentSplit}:

\begin{align}
   (v_{i,z,t})^{2} = (v_{i,z,t}^{re})^{2} + (v_{i,z,t}^{im})^{2} & \label{VoltageSplit}\\
   (i_{i,z,t})^{2} = (i_{i,z,t}^{re})^{2} + (i_{i,z,t}^{im})^{2} & \label{CurrentSplit}
\end{align}

\subsection{Objective function}

The DSO plans an hourly cost-optimal control strategy for the various FRs, to maintain acceptable conditions across its system. The objective (\ref{Objective}) is composed of the following costs: import/export (\ref{IECost}), PV utilization \eqref{PPVCost}-\eqref{QPVCost}, FL alteration (\ref{FLcost}), EV profile re-design (\ref{EVcost}) and limit violation (\ref{SlackCost}):

\begin{align}
\min (C^{I/E} + C^{PPV} + C^{QPV} + C^{FL} + C^{EV} + C^{V}) & \label{Objective}\\
C^{I/E} = c^{I/E} \cdotp \sum_{t, z}\left(P^{I}_{z,t} + P^{E}_{z,t}\right) & \label{IECost}\\
C^{PPV} = c^{PPV} \cdotp \sum_{t, z, p} pf_{p,z,t} \cdotp (S_{p,z,t}^{gen} - S_{p,z,t}^{inv}) & \label{PPVCost}\\
C^{QPV} = c^{QPV} \cdotp \sum_{t, z, p} S_{p,z,t}^{inv} \cdotp \sqrt{1 - pf_{p,z,t}^{2}} & \label{QPVCost}\\
C^{FL} = c^{FL} \cdotp \sum_{t, z, l}(P^{Od,1}_{l,z,t} + P^{Od,2}_{l,z,t} + P^{Ud,1}_{l,z,t} + P^{Ud,2}_{l,z,t})  & \label{FLcost}\\
C^{EV} = \sum_{t, z, e}\left(c^{EV} \cdotp \frac{P^{Oc}_{e,z,t} + P^{Uc}_{e,z,t}}{2} + \xi^{EV} \cdotp c^{DS} \cdotp P^{ds}_{e,z,t}\right)  & \label{EVcost}\\
C^{V} = c^{V} \cdotp \sum_{t, z, i,j: i \neq j}(\sigma_{i,z,t}^{up} + \sigma_{i,z,t}^{down} + \sigma_{ij,z,t}) & \label{SlackCost}
\end{align}
 
Active power imported/exported from/to the MV level incurs a proportional cost for the DSO (\ref{IECost}). PV active power curtailment is highly priced, while the cost for utilizing reactive capabilities is tied to the active power that is not injected due to producing reactive power instead \eqref{PPVCost}-\eqref{QPVCost}. Increasing or decreasing the customer-forecasted load demand has a proportionally associated cost (\ref{FLcost}). Deviations from an EV's customer-desired profile, or discharging,  are proportionally priced (\ref{EVcost}). Technical limits violations carry a high penalty (\ref{SlackCost}). Each cost is chosen based on the desired activation priority order (PO). For example, the less desirable PV curtailment has higher associated costs than load profile alteration. The applied costs (Table \ref{FlexibilityCosts}) simply reflect a certain PO. However, the framework is applicable with any (more realistic) costs.

\begin{table}[b!]
\large
 \centering
\captionof{table}{Characteristics of available FRs and DSO options}
\scalebox{0.64}{
\begin{tabular}{|c|c|c|c|c|}
\midrule
\textbf{FR or DSO option} & \textbf{Service} & \textbf{Type} & \textbf{PO} & \textbf{Cost} (\euro/$kW$)\\ 
\midrule
IE & Energy Import & U & 2 & 1\\
 & Energy Export & D & 2 & 1\\
\hline
PV & Active curtailment & D & 5 & 10\\
 & Reactive management & U/D & 1 & 0.5\\
\hline
FL & Consumption profile & U/D & 3 & 1.5\\
 & alteration & &  & \\
\hline
EV & Charging profile & U/D & 4 & 4.5\\
 & alteration &  & & \\
\hline
Slacks & Limit violations & U/D & 6 & 30\\
\bottomrule
\end{tabular}
}
\caption*{U: Upward, D: Downward}
\label{FlexibilityCosts}
\end{table}

\subsection{Device modelling}

For all subsequent equations, for any device $a$ (load, PV, EV), the symbol $u_{a,z,t}$ refers to \textit{the voltage difference that concerns device $a$, originally connected to phase $z$, at time period $t$}: for 4-wire systems, the difference is between phase and neutral (Wye) or phase (delta); for 3-wire systems, it is between phase and ground (Wye) or phase (delta).

\subsubsection{Customer loads}

The active and reactive demand of a load is based on fixed impedance, current and consumption elements, i.e., the ZIP model (\ref{ActiveDemand})-(\ref{ReactiveDemand}), represented by their respective consumption percentages (\ref{PercentagesSum}). A power factor ($pf$) of 0.95 is assumed for all loads. However, the user may define any $pf$ they deem appropriate. The nominal active and reactive powers, $P^{0}_{l,z,t}, Q^{0}_{l,z,t}$, can thus be interrelated through \eqref{PQrelation}, where $\Omega$ represents the type of $pf$ (1 for lagging, -1 for leading). The load currents are calculated based on (\ref{LoadCurrentReal})-(\ref{LoadCurrentImaginary}). The following hold $\forall l \in {\cal L}, \forall z \in {\cal Z}, \forall t \in {\cal T}$:

\begin{align}
    P^{D}_{l,z,t} = P^{0}_{l,z,t} \left[ a^{Z}_{P} \left( u_{l,z,t} \right)^{2} + a^{I}_{P} \left( u_{l,z,t} \right) + a^{P}_{p,t} \right] & \label{ActiveDemand}\\
    Q^{D}_{l,z,t} = Q^{0}_{l,z,t} \left[ a^{Z}_{Q} \left( u_{l,z,t} \right)^{2} + a^{I}_{Q} \left( u_{l,z,t} \right) + a^{Q}_{p,t} \right] & \label{ReactiveDemand}\\
    a^{Z}_{q,t} + a^{I}_{q,t} + a^{P}_{q,t} = a^{Z}_{p,t} + a^{I}_{p,t} + a^{P}_{p,t} = 1 &\label{PercentagesSum}\\
    Q^{0}_{l,z,t} = P^{0}_{l,z,t} \mid\sin{(\arccos{(pf)})}\mid \cdotp pf^{-1} \cdotp \Omega & \label{PQrelation}\\
    i^{D,re}_{l,z,t} = \left(P^{D}_{l,z,t}u^{re}_{l,z,t} + Q^{D}_{l,z,t}u^{im}_{l,z,t}\right) \cdotp u^{-2}_{l,z,t} & \label{LoadCurrentReal}\\
    i^{D,im}_{l,z,t} =
    \left(P^{D}_{l,z,t}u^{im}_{l,z,t} - Q^{D}_{l,z,t}u^{re}_{l,z,t} \right) \cdotp u^{-2}_{l,z,t} & \label{LoadCurrentImaginary}
\end{align}

A customer load may be flexible (FL) in some ways, in which case its nominal power can be generalized, as shown in \eqref{FlexibilityReplacement}. Highly accurate models split the FL into 4 parts: fixed consumption, $P_{l,z,t}^{fx}$, fully flexible consumption, $P_{l,z,t}^{fc}$, energy (E) shiftable consumption, $P_{l,z,t}^{es}$ and time (T) shiftable consumption $P_{l,z,t}^{ts}$. The first two parts are most commonly considered in the vast majority of cases. The fully flexible consumption is complemented by the variables $P^{Od}_{l,z,t}, P^{Ud}_{l,z,t}$, i.e., ``overdemand" and ``underdemand", respectively. ``Overdemand" means higher consumption than originally forecasted and vice versa for ``underdemand". The two variables are limited by a maximum alteration, $M^{FL}$, and designed to not simultaneously be positive (sub-optimal, see \cite{DOPF}).

While they are less common, the last two parts are also added for the sake of model comprehensiveness. The third part of the FL is energy-shiftable, i.e., in-day alterations are allowed (similarly to the second part), but their net sum must be zero (fixed daily energy consumption), \eqref{EN}. The final part of the FL is time-shiftable, i.e., it can be moved through time, though unaltered. This is modelled through the use of binary variables ($\delta_{l,t}$), see \eqref{TS} which dictates that the fixed part, $P^{SL}_{l,z,t}$, must be active exactly for $A$ time periods (akin to cycle time). The resulting MINLP problem is generally intractable and requires the use of approximations or heuristics to effectively manage, see \cite{SL_Approximations}.

\begin{align}
P_{l,z,t}^{0} \rightarrow P_{l,z,t}^{fx} + P_{l,z,t}^{fc} + P_{l,z,t}^{es} + P_{l,z,t}^{ts} & \label{FlexibilityReplacement}
\end{align}

\begin{equation}
\label{CF}
 \text{Common:}    \left\{ \begin{array}{ll}
            P_{l,z,t}^{fc} \rightarrow  P^{fc}_{l,z,t} + P^{Od,1}_{l,z,t} - P^{Ud,1}_{l,z,t}\\
            0 \leq (P^{Od,1}_{l,z,t}, P^{Ud,1}_{l,z,t}) \leq M^{FL} \cdotp P^{fc}_{l,z,t}
        \end{array} \right.
\end{equation}

\begin{equation}
\label{EN}
  \text{E-shiftable:}    \left\{ \begin{array}{ll}
            P_{l,z,t}^{en} \rightarrow  P^{en}_{l,z,t} + P^{Od,2}_{l,z,t} - P^{Ud,2}_{l,z,t}\\
            0 \leq (P^{Od,2}_{l,z,t}, P^{Ud,2}_{l,z,t}) \leq M^{FL} \cdotp P^{en}_{l,z,t}\\
            \sum_{t}(P^{Od,2}_{l,z,t} - P^{Ud,2}_{l,z,t}) = 0
        \end{array} \right.
\end{equation}
 
\begin{equation}
\label{TS}
 \text{T-shiftable:}   \left\{ \begin{array}{ll}
            P_{l,z,t}^{ts} \rightarrow  P^{ts}_{l,z,t} + P^{SL}_{l,z,t} \cdotp \delta_{l,t}\\
           \sum_{t}(\delta_{l,t}) = A\\
        \end{array} \right.
\end{equation}   

%\textcolor{blue}{With more analytical FL models, the actual gains in terms of objective function improvement are minimal, with respect to the added complexity. While all approaches will be examined, our main focus will be on the first two, i.e., the most common.}

\subsubsection{Photovoltaics}

The apparent power generated by the applied solar irradiation, $S^{gen}_{p,z,t}$, can be curtailed up to a certain percentage, $M^{PV}$ (\ref{PVcurtailment}), resulting into the apparent power at the inverter level, $S^{inv}_{p,z,t}$. The PV's power factor (pf) is partially flexible, allowing for the utilization of a PV's reactive power (\ref{pfDefinition})-(\ref{pfLimits}). Said flexibility is limited by the PV's traditional PCC, as dictated by (\ref{PVPCC}). The injected PV currents are calculated through (\ref{RealPVCurrent})-(\ref{ImaginaryPVCurrent}). When the PV is not connected to a particular phase, all corresponding variables are equal to zero. The following hold $\forall p \in {\cal P}, \forall z \in {\cal Z}, \forall t \in {\cal T}$:

\begin{align}
    S^{gen}_{p,z,t}(1-M^{PV}) \leq S^{inv}_{p,z,t} \leq S^{gen}_{p,z,t} & \label{PVcurtailment}\\
    pf_{p,z,t} = P^{inj}_{p,z,t} \cdotp (S^{inv}_{p,z,t})^{-1} & \label{pfDefinition}\\
    pf_{p,z,t}^{min} \leq pf_{p,z,t} \leq pf_{p,z,t}^{max} & \label{pfLimits}\\
    (Q^{PV}_{p,z,t})^{2} +  (P^{inj}_{p,z,t})^{2} =  (S^{inv}_{p,z,t})^{2}& \label{PVPCC}\\
    P^{inj}_{p,z,t} = u_{p,z,t}^{re}i^{PV,re}_{p,z,t} + u_{p,z,t}^{im}i^{PV,im}_{p,z,t} & \label{RealPVCurrent}\\
    Q^{PV}_{p,z,t} = u_{p,z,t}^{i}i^{PV,re}_{p,z,t} - u_{p,z,t}^{re}i^{PV,im}_{p,z,t} & \label{ImaginaryPVCurrent}    
\end{align}

The above hold for $1\Phi$ PV inverters. However, using 3$\Phi$ inverters instead allows for redistributing a PV's output among the three phases \cite{PhaseBalancing}. For simplicity,  we ignore the PCC (unity $pf$) when phase balancing is possible; nonetheless, per-phase formulations of \eqref{pfDefinition}-\eqref{PVPCC} can be defined. The total injection among the three phases \textit{must be equal} to the original $1\Phi$ injection (\ref{PVOutputMaintenance}). There is a limit to the redistribution of active power, based on the rate of each 1$\Phi$ inverter, $P^{PV,rate}_{p}$, (\ref{PVRedistributionLimit}). An \textit{individual} inverter may consume active power, hence the possible negative values for \textit{1$\Phi$} PV injection. This can be fully exploited at night, when PV production is zero \cite{PhaseBalancing}.

\begin{align}
    \sum_{z}P^{inj}_{p,z,t} = \sum_{z}S^{inv}_{p,z,t} \cdotp pf_{p,z,t} & \label{PVOutputMaintenance}\\
    -P^{PV,rate}_{p}/3 \leq P^{inj}_{p,z,t} \leq P^{PV,rate}_{p}/3 & \label{PVRedistributionLimit} 
\end{align}

\subsubsection{Electric vehicles}

An EV can  ``overcharge" or ``undercharge" with respect to its originally forecasted profile (similar logic as in FLs), or discharge if the V2G capability is available. Since V2G is a much more sought-after service, its cost is much higher than that of profile rescheduling. This design ensures that the ``undercharge" capability (i.e., charging modulation) will be first fully utilized before before discharging can actually be activated (see also \cite{DOPF}). The EV can only interact with the grid when the resident is home (\ref{EVNoWork}). The EV must ``conclude" the optimization horizon under the same status that its owner originally intended \eqref{EVdemand}-\eqref{Positive}, where $n^{c}, n^{d}$ are the charging and discharging efficiencies. Constraint \eqref{EVLimits} dictates the technical limits of the EV's charging/discharging behaviour. The binary parameter $\xi^{EV}$ represents whether the EV can only charge ($\xi^{EV}=0$) or if it has the V2G capability ($\xi^{EV}=1$), in which case it also dabbles as a traditional energy storage system. Constraint \eqref{CL} ensures that the EV does not charge/discharge above/below certain energy limits, $E^{cap}_{e}, E^{min}_{e}$. The EV currents are calculated by (\ref{RealEVCurrent})-(\ref{ImaginaryEVCurrent}). EVs are assumed to operate with a fixed $pf$. Advanced EVs may even be capable of employing 4-quadrant control \cite{EV_4Q}, though such instances are rare; \eqref{pfDefinition}-\eqref{PVPCC} could be easily adapted and employed if need be. When the EV is not connected to a particular phase, all corresponding variables are equal to zero. The following hold $\forall e \in {\cal E}, \forall z \in {\cal Z}, \forall t \in {\cal T}$:

\begin{align}
P^{Oc}_{e,z,t} = P^{Uc}_{e,z,t} = P^{ds}_{e,z,t} = 0 \quad \forall t \in {\cal T}_{e}^{nc}  & \label{EVNoWork}\\
P_{e,z,t}^{demand} = P^{C}_{e,z,t} + P^{Oc}_{e,z,t} - P^{Uc}_{e,z,t} - \xi^{EV} \cdotp P^{ds}_{e,z,t} & \label{EVdemand}\\
(P^{Oc}_{e,z,t}, P^{Uc}_{e,z,t}, P^{ds}_{e,z,t}) \geq 0 & \label{Positive}\\
\sum_{z,t}\left[n^{c} \cdotp (P^{Oc}_{e,z,t} - P^{Uc}_{e,z,t}) - \xi^{EV} \cdotp \frac{P^{ds}_{e,z,t}}{n^{d}} \right] \equiv M = 0 & \label{EVFullCharge}\\
-P^{rate}_{e} \cdotp \xi^{EV} \leq P_{e,z,t}^{demand} \leq P^{rate}_{e}  & \label{EVLimits}\\
E^{min}_{e} \leq M + \sum_{z,t=1}^{t}P^{C}_{e,z,t'} \cdotp n^{c} \leq E^{cap}_{e} - \xi^{EV} \cdotp E^{0}_{e} & \label{CL}\\
P_{e,z,t}^{demand} = u_{e,z,t}^{re}i^{EV,re}_{e,z,t} + u_{e,z,t}^{im}i^{EV,im}_{e,z,t} & \label{RealEVCurrent}\\
    Q^{EV}_{e,z,t} = u_{e,z,t}^{im}i^{EV,re}_{e,z,t} - u_{e,z,t}^{re}i^{EV,im}_{e,z,t} & \label{ImaginaryEVCurrent}  
\end{align}

EVs may also be equipped with $3\Phi$ inverters \cite{PhaseBalancing}. Constraint \eqref{EVLimits} is modified, due to the limit on the redistribution of active power, based on the rate of each $1\Phi$ inverter, $P^{EV,rate}_{e}$, \eqref{EVnewLimits}. While power can be injected in a phase, the EV type is essentially the same (net charger); pure net discharge (V2G) is not possible, as this would involve a different (and far more expensive) kind of EV. In addition, the variables and parameters $P^{C}_{e,t}, P^{Oc}_{e,t}, P^{Uc}_{e,t}$ are now only considered for the EV as a whole (no $z$ index). \eqref{EVdemand} is thus modified as \eqref{newEVdemand}:

\begin{align}
  -P^{EV,rate}_{e}/3 \leq P^{demand}_{e,z,t} \leq P^{EV,rate}_{e}/3 & \label{EVnewLimits}\\ 
 \sum_{z}P_{e,z,t}^{demand} =  P^{C}_{e,t} + P^{Oc}_{e,t} + P^{Uc}_{e,t} & \label{newEVdemand}
\end{align}

\subsubsection{Stand-alone battery}
The EV model is easily adaptable to a stand-alone battery model. This is achievable through the following modifications, where $P_{e,z,t}^{D}$ is the original, user-driven discharging profile, and $P^{Ods}_{e,z,t}, P^{Uds}_{e,z,t}$ are the ``over-discharge" and ``under-discharge" variables, respectively:

\begin{align}
    \eqref{EVdemand}: P_{e,z,t}^{ds} \rightarrow P_{e,z,t}^{D} + P^{Ods}_{e,z,t} - P^{Uds}_{e,z,t}\\
    \eqref{Positive} \rightarrow \eqref{Positive} \, \& \, (P^{Uc} \leq P^{C}_{e,z,t}) \, \& (P^{Uds} \leq P^{D}_{e,z,t})\\
    \eqref{EVFullCharge}: P_{e,z,t}^{ds} \rightarrow P^{Ods}_{e,z,t} - P^{Uds}_{e,z,t}\\
    \eqref{CL}: P^{C}_{e,z,t} \cdotp n^{c} \rightarrow P^{C}_{e,z,t} \cdotp n^{c} - P^{D}_{e,z,t}/n^{d}
\end{align}

\subsection{Low voltage network technical constraints}

\begin{figure}[b!] 
	\centering
	%\vspace{-0.35cm}
	\scalebox{0.3}{\includegraphics[]{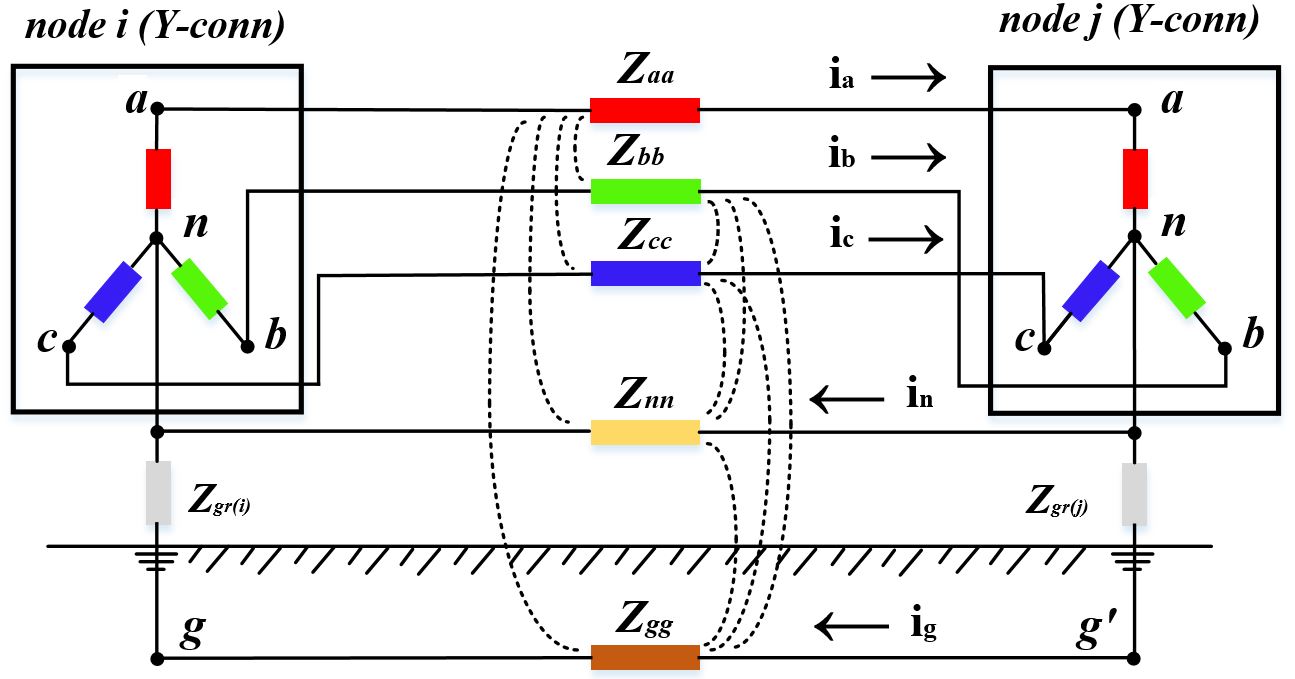}}
	\caption{3$\Phi$, four-wire, multi-grounded distribution line \cite{PF_5phase_Nando}}
	\label{LineModel}
\end{figure}

The distribution line model depicted in Fig. \ref{LineModel} is adopted. Each ``phase" $f \in {\cal F}$ is characterized by its self-impedance, $Z_{ff}$ and its mutual coupling with other ``phases" $\theta \in {\cal F}: f \neq \theta$, $Z_{f\theta}$. A grounding impedance, $Z_{gr}$, may also be present. The neutral and ground currents are calculated using current dividers. As is common for LV networks, devices are assumed to be wye-connected, though the adaptation to delta connections is also presented. For neighboring nodes, currents are assumed equal at origin and destination (small line shunts \cite{PF_5phase_Nando}). 

For the LV network, we have the current injection balance (\ref{CurrentBalance})-(\ref{CurrentBalanceDelta}), where \eqref{CurrentBalance} concerns wye-connected devices and \eqref{CurrentBalanceDelta} details the necessary modification to address delta connections (the superscript $demand$ refers to any considered device, e.g., PV, E, FL, and $z, z^{*}, z^{**}$ represent different phases). The modification is necessary, since employing the conventional $Y \rightarrow \Delta$ transformations would only lead to an approximated MP-OPF solution.  We also have the matching of the injections of the three phases with the neutral phase (\ref{BalanceNeutral}), where binary parameter $\phi^{\Delta}$ represents the existence of a delta connection, the constraints that guarantee the current balance at common ground points (\ref{BalanceGround}), the voltage drop across lines (\ref{RealVoltageDrop})-(\ref{ImaginaryVoltageDrop}), the voltage and branch currents technical limits (\ref{VoltageLimit})-(\ref{CurrentLimit}). The above hold both for the real and imaginary parts. Note that all slack variables are positive (\ref{SlackLimit}). No constraints are enforced for the neutral and ground ``phases" since they are completely dependent on phases a, b, c. The balances between the currents at the neutral-ground connection are enforced by the current dividers; no additional constraints are necessary to capture this behavior. The following hold $\forall i,j \in {\cal I}: i \neq j, \forall z \in {\cal Z}, \forall f \in {\cal F}, \forall t \in {\cal T}$:

\begin{align}
 Y: \, - i^{PV,re/im}_{i,z,t} + i^{D,re/im}_{i,z,t} + i^{EV,re/im}_{i,z,t} = \sum i^{re/im}_{ij,z,t} & \label{CurrentBalance}\\
 \Delta: \, i^{device, re/im}_{i,z,t} \rightarrow i^{device, re/im}_{i,z-z^{*},t} + i^{device, re/im}_{i,z-z^{**},t} & \label{CurrentBalanceDelta}\\
 \sum_{z}i^{re}_{ij,z,t} = i^{re}_{i,n,t} \cdotp \phi^{\Delta} \quad \& \quad \sum_{z}i^{im}_{ij,z,t} = i^{im}_{i,n,t} \cdotp \phi^{\Delta}  \label{BalanceNeutral}\\
  \sum i^{re}_{ij,g,t} = i^{re}_{i,g,t} \quad \& \quad \sum i^{im}_{ij,g,t} = i^{im}_{i,g,t} & \label{BalanceGround}\\
 u_{i,f,t}^{re} - u_{j,f,t}^{re} =\sum_{\theta \in {\cal F}} \left (R_{l, f\theta} i^{re}_{l,\theta} - X_{l, f\theta} i^{im}_{l,\theta}\right ) & \label{RealVoltageDrop}\\
 u_{i,f,t}^{im} - u_{j,f,t}^{im} =\sum_{\theta \in {\cal F}} \left (R_{l, f\theta} i^{im}_{l,\theta} + X_{l, f\theta} i^{re}_{l,\theta}\right ) & \label{ImaginaryVoltageDrop}\\
 V^{min} - \sigma^{down}_{i,z,t} \leq u_{i,z,t} \leq V^{max} + \sigma^{up}_{i,z,t} & \label{VoltageLimit}\\
 -I^{max} - \sigma_{ij,z,t} \leq i_{ij,z,t} \leq I^{max} + \sigma_{ij,z,t} & \label{CurrentLimit}\\
 \sigma_{i,z,t}^{up}, \sigma_{i,z,t}^{down}, \sigma_{ij,z,t} \geq 0 & \label{SlackLimit}
\end{align}

\subsection{Voltage shifts}

\begin{figure}[b!] 
	\centering
	%\vspace{-0.35cm}
	\scalebox{0.33}{\includegraphics[]{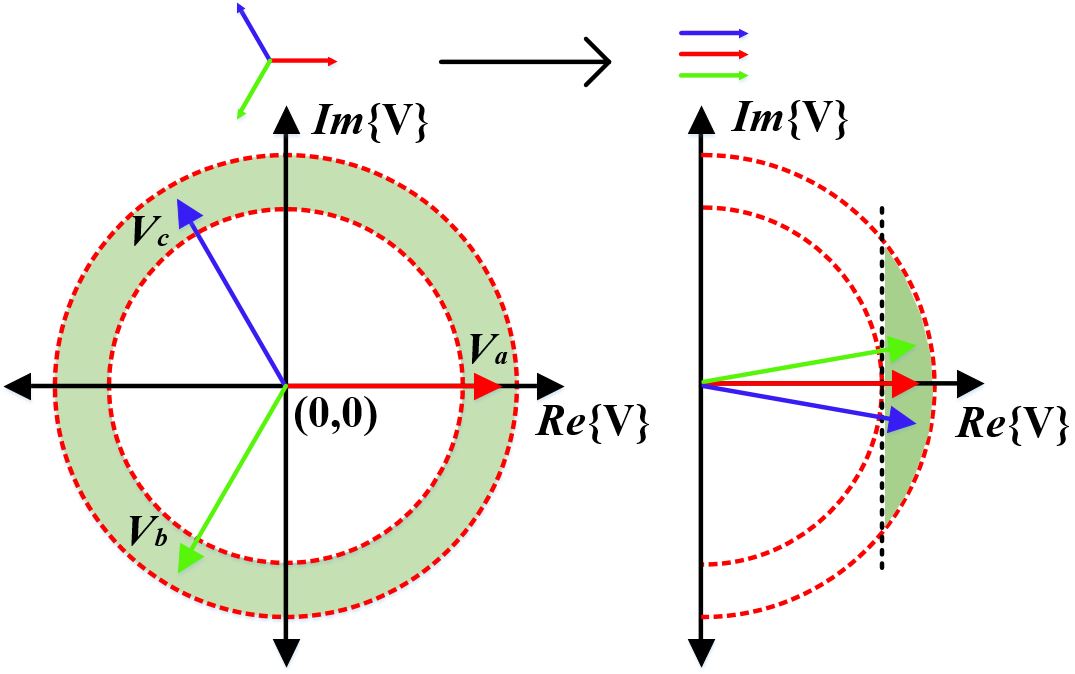}}
	\caption{Voltage constraint reformulation (adapted from \cite{Karagiannopoulos_PSCC})}
	\label{VoltageShift}
\end{figure}

At the feeder head, voltages are defined as \{$1\angle 0^{^\circ}, 1\angle -120^{^\circ}, 1\angle 120^{^\circ}$\} $\rightarrow$ \{$1+0j, -0.5-0.87j, -0.5+0.87j$\} for phases a, b, c, respectively. While constraint \eqref{VoltageLimit} is perfectly valid, a computational inefficiency stems from the fact that the phases are rotated with respect to each other; the rectangular modelling of voltages causes constraint (\ref{VoltageLimit}) to be non-convex. Assuming that the voltage angles (per phase) diverge only slightly from their reference value, a convex reformulation is employed, similarly to \cite{Karagiannopoulos_PSCC}. This allows overcoming one of the several non-convexities, while making the evaluation of the results far easier and more intuitive. The three phases are rotated by ${\cal ROT} = -\{1\angle 0^{\circ}, 1\angle -120^{\circ}, 1\angle 120^{^\circ}\}$ so that they lie close to the reference axis $0^{\circ}$, and the same feasible space is defined for each \cite{Karagiannopoulos_PSCC}. A visual representation is presented in Fig. \ref{VoltageShift}. Constraint (\ref{VoltageLimit}) is thus re-defined as:

 \begin{equation}
\label{VoltageLimitNew}
\left\{ \begin{array}{ll}
  {\cal ROT}(u_{i,z,t}) \leq V^{max} + \sigma_{i,z,t}^{up}\\
 Re\{{\cal ROT}(u_{i,z,t})\} \geq V^{min} - \sigma^{down}_{i,z,t}\}
\end{array} \right.    
\end{equation}

A crucial point not addressed in \cite{Karagiannopoulos_PSCC} (as the paper assumed a perfectly grounded system and consequently used the Kron reduction to remove the neutral cable) is that the phase rotation affects the interactions between phases a, b, c and the neutral wire (\ref{BalanceNeutral}). These constraints must be updated to account for the voltage rotation performed in (\ref{VoltageLimitNew}). As such, the calculated current for each phase is counter-rotated by ${\cal ROT}$. If we define the total current injections at node bus $i$, phase $z$, period $t$, (\ref{InjectionRepresentation}), then (\ref{BalanceNeutral}) are updated as (\ref{RealBalanceNeutralNew})-(\ref{ImagninaryBalanceNeutralNew}):

\begin{align}
     -\sum i^{re}_{ij,z,t} = y^{re}_{i,z,t} \quad \& \quad -\sum i^{im}_{ij,z,t} = y^{im}_{i,z,t} & \label{InjectionRepresentation}
\end{align}
\vspace{-8mm}     

\begin{equation}
\label{RealBalanceNeutralNew}
\begin{split}
y^{re}_{i,a,t} &- (0.5y^{re}_{i,b,t} + 0.87y^{im}_{i,b,t}) \\
&- (0.5y^{re}_{i,c,t} - 0.87y^{im}_{i,c,t}) = -i^{re}_{i,n,t}   
\end{split}  
\end{equation} 
\begin{equation}
\label{ImagninaryBalanceNeutralNew}
\begin{split}
y^{im}_{i,a,t} &- (0.5y^{im}_{i,b,t} - 0.87y^{re}_{i,b,t}) \\
&- (0.5y^{im}_{i,c,t} + 0.87y^{re}_{i,c,t}) = -i^{im}_{i,n,t}    
\end{split}  
\end{equation}
      
\subsection{Remarks}

The problem at hand is an MP-OPF. The conventional setup (1$\Phi$ inverters, no voltage shifting) is comprised of the rectangular coordinates modelling and the various costs \eqref{VoltageSplit}-\eqref{SlackCost}, the modelling of FLs \eqref{ActiveDemand}-\eqref{TS}, PVs \eqref{PVcurtailment}-\eqref{ImaginaryPVCurrent}, EVs \eqref{EVNoWork}-\eqref{ImaginaryEVCurrent}, and the system technical constraints \eqref{CurrentBalance}-\eqref{SlackLimit}. Proposed novel aspects of this MP-OPF pertain to goal (1) are: ZIP load flexibility, realistic PV power management, EV original profile alteration and neutral/ground currents interactions. 

Further novelty allows adopting an advanced MP-OPF formulations. PVs are additionally complimented by the specific constraints of their 3$\Phi$ inverters \eqref{PVOutputMaintenance}-\eqref{PVRedistributionLimit}. Same goes for EVs, where the original total demand constraints \eqref{EVLimits} are replaced by the updated \eqref{EVnewLimits}-\eqref{newEVdemand}. Voltage constraint \eqref{VoltageLimit} is replaced by the set of shifted constraints \eqref{VoltageLimitNew}. The original relations between the currents of phases a, b, c and the neutral current \eqref{BalanceNeutral} are converted to reflect the voltage shift \eqref{InjectionRepresentation}. 

The physical problem modelled is an NLP, regardless of the modelling choices concerning the network and the residential devices. The nonlinearity stems from the ZIP load models, the PCC of PVs and the relations between power and voltage/current. The complex formulation, including all the peculiarities of MP-OPF in multi-phase LV networks, is not directly amenable to exact convex formulations (particularly SDP/SOCP). The authors prefer adopting an NLP formulation and solver, in order to obtain a feasible and at least local optimal solution, as compared to the (high) risk of obtaining physically meaningless solution from a relaxed problem \cite{ACOPF_DS}.

\section{Case Studies and Framework Exhibition}

\subsection{Simulation environment}

\begin{figure}[b!] 
	\centering
	%\vspace{-0.35cm}
	\scalebox{0.33}{\includegraphics[]{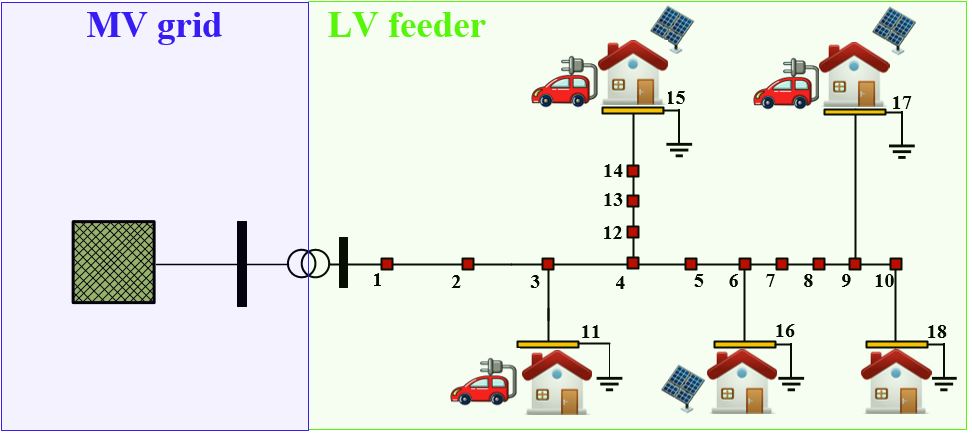}}
	\caption{CIGRE LV feeder (28\% customer-to-node ratio)}
	\label{Network}
\end{figure}

The proposed formulation is applied on the 18-node, modified CIGRE LV benchmark distribution network \cite{CigreLV}, depicted in Fig. \ref{Network}. The original network dataset does not include data regarding the neutral wire and the ground; artificial values are used, based on observations regarding the relationship between phases a, b, c and the missing values, as derived from \cite{PF_5phase_Nando}. When grounding is included, a value of $R_{g} = 1 \Omega$ (adjusted in p.u.) is considered for all customer nodes, though other kinds of grounding are also possible (e.g., grounding impedance or Petersen coil). All other nodes are assumed to be ungrounded. Base power and (3$\Phi$) voltage are chosen as $1 kW$ and $240 V$, respectively. The Z/I/P percentages are randomly assigned.

The connected distributed devices throughout the network are FLs, EVs and PVs (connection points shown in Fig. \ref{Network}). The characteristics of each device are available in Table \ref{T5}. Their actual 24-hour profiles are available in \cite{DSF_AVRA}. While the framework can accommodate any residential device, in order to better examine the impact of different flexibility setups, devices with high power rates are purposefully selected. All simulations are performed on a PC of 2.7-GHz and 8-GB RAM, using the general purpose NLP solver IPOPT \cite{IPOPT}, with default settings, through GAMS \cite{GAMS}. The framework was previously evaluated (via power flow comparisons) on a number of MV and LV networks. The errors (in p.u.) on complex voltages varied between 10$^{-4}$ and 10$^{-6}$.

\begin{table}[b!]
\large
 \centering
\captionof{table}{Device characteristics}
\scalebox{0.6}{
\begin{tabular}{|c|c|c|c|c|c|c|c|}
\midrule
\textbf{Node} & \textbf{Device} & $P^{r} (kW)$ & $E^{cap} (kWh)$ & \textbf{Phase} & $a^{Z}_{P/Q}$ & $a^{I}_{P/Q}$ & $a^{P}_{P/Q}$\\ 
\midrule
11 & EV & 8 & 24 & a & -- & -- & --\\
\hline
11 & FL & 1.5 & -- & a & 0.2/0.1 & 0.2/0.1 & 0.6/0.8\\
\hline
15 & EV & 8  &  24 & b & -- & -- & --\\
\hline
15 & FL & 2  &  -- & b & 0.6/0.3 & 0.1/0.2 & 0.3/0.5\\
\hline
15 & PV & 4  &  -- & a & -- & -- & --\\
\hline
16 & FL & 2.5  & -- & c & 0.05/0.3 & 0.15/0.1 & 0.8/0.6\\
\hline
16 & PV & 4  &  -- & b & -- & -- & --\\
\hline
17 & EV & 10  &  30 & c & -- & -- & --\\
\hline
17 & FL & 1.5 & -- & a & 0.3/0.5 & 0.4/0.1 & 0.3/0.4\\
\hline
17 & PV & 5  & -- & c & -- & -- & --\\
\hline
18 & FL & 2  & -- & b & 0.05/0.01 & 0.25/0.8 & 0.7/0.1\\
\bottomrule
\end{tabular}
}
\label{T5}
\end{table}

\subsection{Modelling and operational scenarios}

Apart from proposing a comprehensive MP-OPF framework for realistic distribution systems, this work also provides information on the impact of each modelling aspect on the solution. The goal is to understand which are necessary and which can be reasonably ignored, depending on the needs of the application. Three major, multi-scenario cases are examined:

\begin{table}[b!]
\large
 \centering
\captionof{table}{Versions of load modelling}
\scalebox{0.7}{
\begin{tabular}{|c|c|c|c|c|c|c|c|}
\midrule
Version & $L_{1}$ & $L_{2}$ & $L_{3}$ & $L_{4}$ & $L_{5}$ & $L_{6}$ & $L_{7}$\\ 
\midrule
Z-component & \checkmark & \checkmark & \checkmark & -- & \checkmark & -- & --\\
\hline
I-component & \checkmark & \checkmark & -- & \checkmark & -- & \checkmark & --\\
\hline
P-component & \checkmark & -- & \checkmark & \checkmark & -- & -- & \checkmark\\
\hline
VD & Q \& LIN & Q \& LIN & Q & LIN & Q & LIN & None\\
\bottomrule
\end{tabular}
}
\caption*{Q: Quadratic, LIN: Linear, VD: Voltage dependency}
\label{T1}
\end{table}

\begin{table}[b!]
\large
 \centering
\captionof{table}{Versions of network modelling}
\scalebox{0.7}{
\begin{tabular}{|c|c|c|c|c|}
\midrule
Version & Grounding & Neutral & Mutual & Unbalanced\\ 
 & &  & Coupling & \\
\midrule
$N_{1}$ (3$\Phi$, 4-wire) & \checkmark & \checkmark & \checkmark & \checkmark\\
\hline
$N_{2}$ (3$\Phi$, 4-wire) & -- & \checkmark & \checkmark & \checkmark\\
\hline
$N_{3}$ (3$\Phi$, 3-wire) & -- & -- & \checkmark & \checkmark\\
\hline
$N_{4}$ (3$\Phi$, 3-wire) & -- & -- & -- & \checkmark\\
\hline
$N_{5}$ (1$\Phi$) & -- & -- & -- & --\\
\bottomrule
\end{tabular}
}
%\caption*{Dev: Deviation}
\label{T2}
\end{table}

\begin{table}[b!]
\large
 \centering
\captionof{table}{FR controllability scenarios}
\scalebox{0.7}{
\begin{tabular}{|c|c|c|c|c|c|c|c|c|}
\midrule
Scenario & $S_{1}$ & $S_{2}$ & $S_{3}$ & $S_{4}$ & $S_{5}$ & $S_{6}$ & $S_{7}$ & $S_{8}$\\
\midrule
$M^{PV}$ & 0 & 0 & 0 & 0.2 & 0.2 & 0.2 & 0.2 & 0.2\\
\hline
PV $pf^{min}$ & 1 & 0.9 & 0.8 & 0.7 & 0.7 & 0.7 & 0.7 & 0.7\\
\hline
$M^{FL}$ & 0 & 0 & 0 & 0 & 0.1 & 0.3 & 0.3 & 0.3\\
\hline
EV reschedule & -- & -- & -- & -- & -- & -- & \checkmark & \checkmark \\
\hline
${\cal T}^{nc}$ & -- & -- & -- & -- & -- & -- & $t_{4}-t_{17}$ & $t_{8}-t_{15}$\\
\bottomrule
\end{tabular}
}
%\caption*{Dev: Deviation}
\label{T4}
\end{table}

\begin{enumerate}
    \item Case 1: Unbalanced network model (3$\Phi$, perfectly grounded neutral), $1\Phi$ PVs/EVs ($\xi^{EV}=0$). Various load models, $L_{x}$, are examined; they are presented in Table \ref{T1}. PVs can be curtailed up to 20\% ($M^{PV} = 0.2$), FLs modified up to 10\% ($M^{FL} = 0.1$). 
    \item Case 2: ZIP model for customer loads, $1\Phi$ PVs/EVs ($\xi^{EV}=0$). Various network models, $N_{x}$, are examined; they are presented in Table \ref{T2}. PVs can be curtailed up to 20\%, FLs modified up to 10\%. 
    \item Case 3: Unbalanced network model (3$\Phi$ plus neutral) and ZIP model for customer loads. Two scenarios of inverter types are examined: 1$\Phi$ inverters vs 3$\Phi$ inverters with balancing capabilities ($\xi^{EV}=0$). Various scenarios of FR controllability are examined; the different scenarios $S_{x}$ are presented in Table \ref{T4}. 
\end{enumerate}

Focusing on smart distribution feeders with high penetration of potential FRs, the main purpose of this work is reflected in Cases 1 and 2. That purpose is to provide a comprehensive understanding to DSOs with regards to how their system realistically behaves, how the accuracy of the solution is affected with cumulative simplifications/approximations, and how practical each level of modelling detail actually is. The inclusion of Case 3 is done in order to demonstrate to DSOs how realistic distribution systems would behave if the full potential of the various interconnected FRs were unlocked. 

\begin{table*}[t!]
\large
 \centering
\captionof{table}{Results per load model: 500 Monte Carlo simulations per model with varying ($\pm$ 10\%) load/PV profiles}
\scalebox{0.67}{
\begin{tabular}{|c|c|c|c|c|c|c|c|}
\toprule
& $L_{1}$ & $L_{2}$ & $L_{3}$ & $L_{4}$ & $L_{5}$ & $L_{6}$ & $L_{7}$\\
 \midrule
Solution time (s) & 4.59 $\pm$ 1.19 & 6.52 $\pm$ 5.2 & 5.68 $\pm$ 5.04 & 3.83 $\pm$ 4.25 & 5.58 $\pm$ 0.55 & 5.22 $\pm$ 2.30 & 5.78 $\pm$ 0.83  \\
\hline
Objective function (\euro) & 340.8 $\pm$ 1.9 & 330.5 $\pm$ 1.8 & 319.6 $\pm$ 1.7 & 309.8 $\pm$ 1.6 & 323.7 $\pm$ 1.7 & 338.7 $\pm$ 1.6 & 357.3 $\pm$ 2.0  \\
\hline
Average system voltage (p.u.) & 0.982 $\pm$ 0.01 & 0.983 $\pm$ 0.02 & 0.983 $\pm$ 0.02 & 0.981 $\pm$ 0.01 & 0.983 $\pm$ 0.02 &  0.982 $\pm$ 0.01 & 0.980 $\pm$ 0.03 \\
\hline
Cumulative limit violations (p.u.) & 1.65 $\pm$ 0.02 & 1.55 $\pm$ 0.03 & 1.48 $\pm$ 0.02 & 1.27 $\pm$ 0.02 & 1.47 $\pm$ 0.03 & 1.67 $\pm$ 0.03 & 1.86 $\pm$ 0.04 \\
\hline
Average customer demand (p.u.) & 2.31 $\pm$ 0.14 & 2.26 $\pm$ 0.13 & 2.07 $\pm$ 0.15 & 2.12 $\pm$ 0.20 & 2.21 $\pm$ 0.13 & 2.28 $\pm$ 0.13 & 2.40 $\pm$ 0.14  \\
\hline
 \multicolumn{8}{|c|}{Approximate cost breakdown: IE (77.9 \%), FLs (1.76 \%), PVs (2.35 \%), EVs (3.23 \%), Slacks (14.71 \%)}\\
\bottomrule
\end{tabular}
}
%\caption*{37,653 variables, 17,001 constraints}
\label{LoadResults}
\end{table*}

 \begin{table*}[t!]
\large
 \centering
\captionof{table}{Results per network model: 500 Monte Carlo simulation per model with varying ($\pm$ 10\%) load/PV profiles}
\scalebox{0.7}{
\begin{tabular}{|c|c|c|c|c|c|}
\toprule
& $N_{1}$ & $N_{2}$ & $N_{3}$ & $N_{4}$ & $N_{5}$\\
 \midrule
Solution time (s) & 17.45 $\pm$ 5.32 & 12.81 $\pm$ 6.41 & 5.37 $\pm$ 1.73 & 8.39 $\pm$ 3.80 & 6.83 $\pm$ 4.21   \\
\hline
Objective function (\euro) & 397.0 $\pm$ 2.1 & 392.2 $\pm$ 1.9 & 332.9 $\pm$ 1.8 & 275.9 $\pm$ 1.6 & 291.9 $\pm$ 2.0  \\
\hline
Average system voltage (p.u.) & 0.981 $\pm$ 0.05 & 0.980 $\pm$ 0.03 & 0.982 $\pm$ 0.02 & 0.985 $\pm$ 0.05 & 0.935 $\pm$ 0.07  \\
\hline
Cumulative limit violations (p.u.) & 1.46 $\pm$ 0.05 & 1.44 $\pm$ 0.03 & 1.66 $\pm$ 0.03 & 0.39 $\pm$ 0.03 & 0.65 $\pm$ 0.01 \\
\hline
Average customer demand (p.u.) & 2.41 $\pm$ 0.12 & 2.42 $\pm$ 0.14 & 2.35 $\pm$ 0.19 & 2.51 $\pm$ 0.17 & 2.55 $\pm$ 0.22\\
\hline
Total variables & 53,097 & 43,569 & 37,653 & 37,653 & 12,885 \\
\hline
Total constraints & 24,681 & 22,953 & 17,001 & 17,001 &  3,848\\
\bottomrule
\end{tabular}
}
%\caption*{$1\Phi$: Single-phase inverters, $3\Phi$: 3-phase inverters}
\label{NetworkResults}
\end{table*}

\subsection{Examining the effects of modelling choices}

\subsubsection{Case 1}

Results for various levels of load modelling accuracy (problem of 37,653 variables, 17,001 constraints) are presented in Table \ref{LoadResults}. It becomes immediately apparent that the (often neglected) load modelling choices affect the solution in a non-negligible way (though solution times and total violations are comparable). When the Z-component is dominant, the network's voltage profile tends to rise. Contrariwise, when the P-component is dominant, the voltage profile tends to drop. The I-component appears to have the smallest impact on the voltage profile. Similar observations are much more prevalent in nodes that are further away from the substation. 

A higher average demand does not necessarily correspond to a reduced voltage profile. When there are binding interactions between the load model and the voltage profile, it is not always straightforward to estimate how the two will be affected. Formulations that include the P-component (higher average demand) such as ZP, IP, P, tend to show pictures of higher stress for the system, requiring more reactionary measures in response (i.e., higher operational costs). Formulations that can produce a more ``malleable" demand such as ZI, Z, I, have the opposite effect (lower costs on average). The smallest, almost negligible difference in the objective is observed under the ZP load model; the I-aspect of loads could perhaps safely be dropped from simplified formulations. Nevertheless, less conservative load models (assuming there is sufficient data to support them) can bring about operational benefits (more accurate coordination of flexibility options). Do note that while the differences (not necessarily errors) in objective values are not very high (on average about 3.3\%), the cumulative impact of the load modelling decisions are certainly non-negligible.

\subsubsection{Case 2}

Results for various levels of network modelling accuracy are presented in Table \ref{NetworkResults} (assuming $N_{1}$ is the ideal objective). Simpler network models produce inaccuracies in the voltage profile (elevated or reduced levels), leading to significant overestimations or underestimations of operational issues (ignoring imbalances, mutual coupling or both produces an average error of 29.25\%). Models $N_{4}, N_{5}$ have unrealistically low violation costs, despite exhibiting vastly different voltage profiles. While simpler models have benefits (speed-wise), they could be inappropriate for many applications.

When the neutral wire is explicitly modelled the error drops to approximately 1.25\%. The large errors observed are not unexpected. As was observed in \cite{OptimalBSSDispatchUnbalanced}, load imbalances of about 50\% can lead to the corresponding neutral wire being responsible for approximately 20\% of system losses. In our case, some nodes have 100\% imbalance, with the neutral wire carrying very high currents. While the neutral voltages are very low (almost always below 0.1 p.u.), the coupling effects of the large neutral currents create several issues. The Kron reduction is acceptable only on systems with perfectly grounded neutral wires; this is not our case, hence we observe large errors. Depending on the level of imbalance, the modelling aspect that could be reliably dropped is the grounding.  

\begin{table*}[t!]
\large
 \centering
\captionof{table}{Results, 1$\Phi$ inverters vs 3$\Phi$ inverters}
\scalebox{0.65}{
\begin{tabular}{|c|c|c|c|c|c|c|c|c|c|c|c|c|c|c|c|c|c|c|c|c|}
\toprule
& \multicolumn{2}{|c|}{$S_{1}$} & \multicolumn{2}{|c|}{$S_{2}$} & \multicolumn{2}{|c|}{$S_{3}$} & \multicolumn{2}{|c|}{$S_{4}$}
& \multicolumn{2}{|c|}{$S_{5}$} & \multicolumn{2}{|c|}{$S_{6}$} & \multicolumn{2}{|c|}{$S_{7}$} & \multicolumn{2}{|c|}{$S_{8}$}\\
\midrule
& $1\Phi$ & $3\Phi$ & $1\Phi$ & $3\Phi$ & $1\Phi$ & $3\Phi$ & $1\Phi$ & $3\Phi$ & $1\Phi$ & $3\Phi$ & $1\Phi$ & $3\Phi$ & $1\Phi$ & $3\Phi^{*}$ & $1\Phi$ & $3\Phi^{*}$ \\
 \midrule
 Solution time (s) & 132.03 & 29.30 & 56.06 & 29.30 & 67.73 & 29.30 & 5.94 & 29.30 & 4.92 & 22.54 & 4.75 & 26.63 & 7.20 & 22.05 & 26.77 & 21.11  \\
 \hline
 Objective function (\euro) & 465 & 2,663 & 459 & 2,663 & 452 & 2,663 & 446 & 2,663 & 442 & 2,527 & 433 & 2,215 & 433 & 2,258 & 416 & 2,241  \\
 \hline
 Average system voltage (p.u.) & 0.981 & 0.982 & 0.982 & 0.982 & 0.978 & 0.982 & 0.981 & 0.982 & 0.982 & 0.981 & 0.982 & 0.982 & 0.982 & 0.982 & 0.982 & 0.983  \\
\hline
Maximum voltage (p.u.) & 1.083 & 1.067 & 1.1 & 1.067 & 1.091 &  1.067 & 1.098 & 1.067 & 1.098 & 1.075 & 1.098 & 1.061 & 1.098 & 1.071 & 1.098 & 1.055 \\
\hline
Minimum voltage (p.u.) & 0.798 & 0.902 & 0.798 & 0.902 & 0.798 &  0.902 & 0.798 & 0.902 & 0.802 & 0.902 & 0.809 & 0.903 & 0.809 & 0.908 & 0.809 & 0.908 \\
\bottomrule
\end{tabular}
}
\caption*{3$\Phi$ case: PV $pf=1$, IE price increased tenfold to motivate system self-sufficiency. *Lower voltage limit set to 0.95 p.u.} 
\label{InverterResults}
\end{table*}

\subsubsection{Case 3}

Results for various levels of FR controllability, using both $1\Phi$ and $3\Phi$ inverters, are presented in Table \ref{InverterResults}. Expectedly (for the $1\Phi$ case), as more flexibility is added to the DSO's ``toolkit" the operational conditions improve. In our case, increased PV controllability is followed by increased FL controllability, finally adding EV controllability into the mix. The voltage profile remains relatively steady, though the upper and lower values are improved, resulting in reduced violations costs. In addition, the available residential flexibility also reduces the import costs, as an increasing amount of in-system energy is re-distributed, decreased or converted (e.g., PVs changing their $pf$ to avoid curtailment).

For the $3\Phi$ inverters case the IE price is increased 10 times in order to motivate system self-sufficiency, discourage (as much as possible) interactions with the MV level and maximize the utility of residential FRs. Even without employing any flexibility, the usage of $3\Phi$ inverters improves the voltage profile; even for $S_{1}$, we observe total violations of almost zero. The usage of $3\Phi$ inverters offers tremendous amounts of flexibility to the operator (despite the fact that we have also removed the reactive capabilities of PVs), without the need to actually engage in profile alterations. A thing of note is that when the connected devices are able to distribute their profiles between the three phases, the neutral wires carry almost no current and as such most of the original active losses are no longer observed; had the original IE cost been kept then the objective function value (as compared to the $1\Phi$ inverters case) \textit{would have been reduced by about 55\%}.

\subsection{Assessing scalability}

\subsubsection{System vs customer base expansion}

Similarly to most NLP problems, full scalability for very large problems is not guaranteed. Nonetheless, the formulation is in general computationally efficient for small/medium-size systems and sometimes even for large systems (see Section III.D.2). For the 18-node system, despite the already large problem size and the variety of available flexibility options, a solution is achieved (on average) in less than 30 seconds. 

\begin{figure}[b!] 
	\centering
	\vspace{-0.35cm}
	\scalebox{0.16}{\includegraphics[]{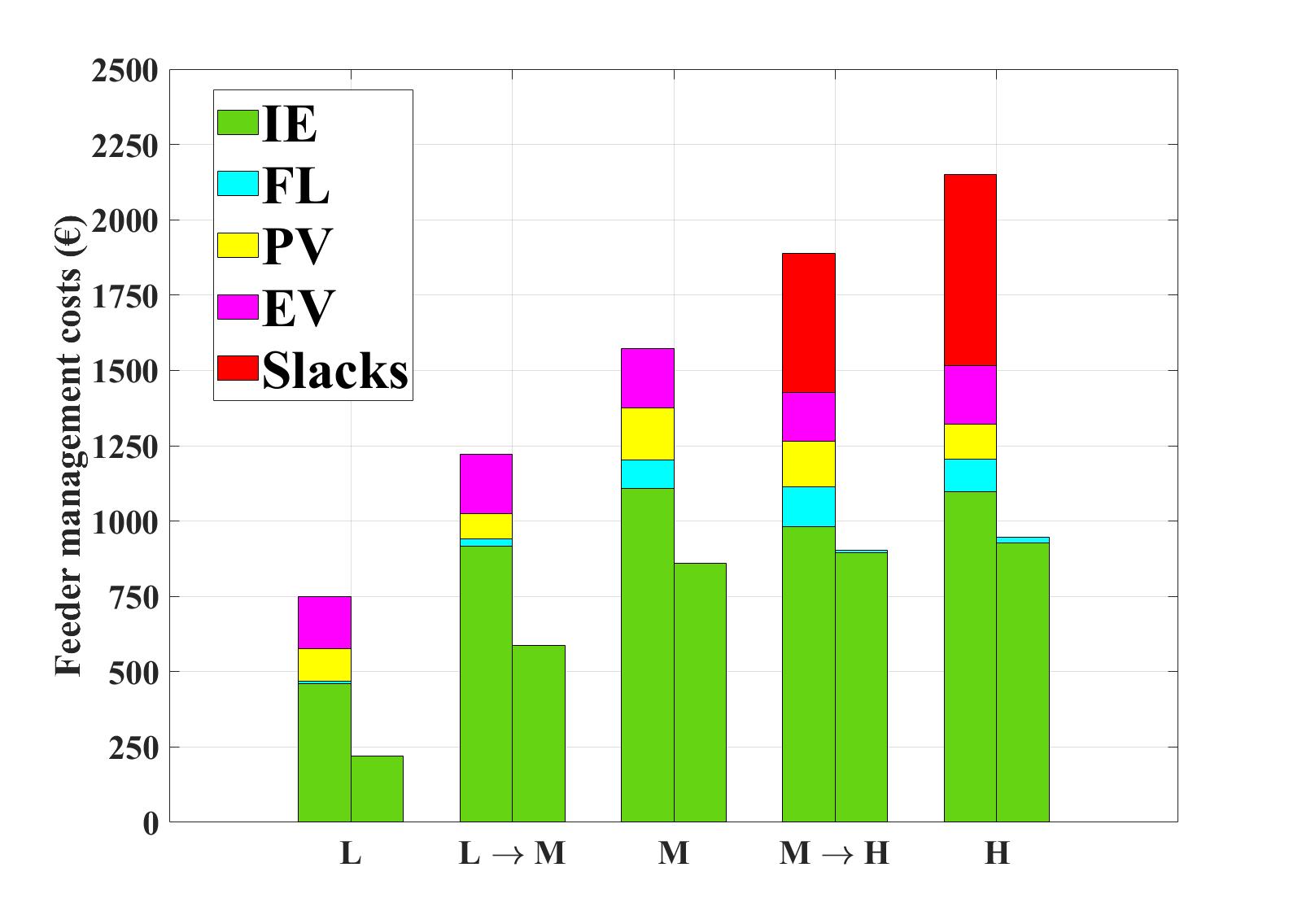}}
	\caption{Average cost breakdown for 180-node system: 1$\Phi$ (left) vs 3$\Phi$ (right) inverter setups}
	\label{180Results}
\end{figure}

The proposed formulation is applied on a partial real-life distribution feeder, composed of 180 nodes hosting 23 residential customers (13\% customer-to-node ratio, see \cite{DSF_AVRA} for original), of which 5 own EVs, 8 own PVs and 10 own both (sizes of Table \ref{T5}). The system is assumed perfectly grounded and as such, the neutral wire can be reliably reduced. Five different loading conditions scenarios are examined, using both the 1$\Phi$ and 3$\Phi$ inverter setups: light (L, one third of the regular consumption), medium (M, regular consumption), heavy (H, triple the regular consumption) and in-between states. These conditions were examined for the sake of evaluating the flexibility potential of FRs under extremely challenging conditions. The results are presented in Fig. \ref{180Results} and Table \ref{T180}.

\begin{table}[b!]
\large
 \centering
\captionof{table}{Solution performance, 180-node system: 50 Monte Carlo simulations per loading level with varying ($\pm$ 10\%) load/PV profiles.}
\scalebox{0.65}{
\begin{tabular}{|c|c|c|c|c|}
\midrule
& \multicolumn{4}{c}{Metric}\\
Loading & System voltage  & System voltage & Solution time & Solution time\\ 
level & (pu, 1$\Phi$) & (pu, 3$\Phi$) & (s, 1$\Phi$) & (s, 3$\Phi$)\\
\midrule
L & 0.994 $\pm$ 0.03 & 0.998 $\pm$ 0.02 & 250 $\pm$ 11 & 230 $\pm$ 5 \\ 
\hline
L $\rightarrow$ M  & 0.978 $\pm$ 0.08 & 0.985 $\pm$ 0.06 & 472 $\pm$ 16 & 248 $\pm$ 7\\
\hline
M & 0.961 $\pm$ 0.12 & 0.972 $\pm$ 0.09 & 766 $\pm$ 24 & 267 $\pm$ 13 \\
\hline
M $\rightarrow$ H & 0.944 $\pm$ 0.16 & 0.957 $\pm$ 0.13 & 1558 $\pm$ 69 & 301 $\pm$ 20 \\
\hline
H & 0.923 $\pm$ 0.2 & 0.942 $\pm$ 0.17 & 1920 $\pm$ 124 & 329 $\pm$ 37 \\
\bottomrule
\end{tabular}
}
%\caption*{Q: Quadratic, LIN: Linear}
\label{T180}
\end{table}

The results illustrate some interesting points. While the problems sizes are comparable \footnote{Remember that for loads, PVs and EVs, the variables of unconnected phases are still present, despite being set to zero.}, the 3$\Phi$ setup has the superior performance. After increasing the degree of stressing of the system per step (indicated by the increasing objective value), an optimal solution without violations is always achieved, with the solution speed increasing, but not significantly. Contrariwise, the 1$\Phi$ setup cannot cope efficiently with high levels of stress, requiring more time to reach a solution (on average 270\% slower). In fact, to avoid infeasibility for the H case, much more residential flexibility is ``activated" than would be reasonably expected. The 3$\Phi$ setup provides far better solutions (on average 54.4\% better), due to not directly utilizing flexibility per se but rather by  alleviating the negative effects of load imbalances through consumption redistribution amongst phases. The above are indicative of the superiority of ``investing" in flexibility that is characterized by quality (phase balancing) instead of quantity (profile alteration).

\begin{table}[t!]
\large
 \centering
\captionof{table}{Solution performance, 180-node system, significantly expanded customer base}
\scalebox{0.65}{
\begin{tabular}{|c|c|c|}
\midrule
& 3$\Phi$ inverters & 1$\Phi$ inverters\\ 
\midrule
Solution time (s) & 49.6 $\pm$ 8.1 & 127.2 $\pm$ 18.7\\
\hline
Objective function (\euro) & 4,220 $\pm$ 340 & 9,500 $\pm$ 610\\
\hline
Cumulative limit violations (p.u.) & 0 $\pm$ 0 & 0.2 $\pm$ 0.05\\
\hline
``Activated" FLs (\%) & 2 $\pm$ 0.7 & 55 $\pm$ 7\\
\hline
EV ``activation"/phase balancing (\%) & 0/90 $\pm$ 0/1.5 & 87/0 $\pm$ 4/0\\
\hline
PV ``activation"/phase balancing (\%) & 0/95 $\pm$ 0/2 & 70/0 $\pm$ 8/0\\
\hline
 \multicolumn{3}{|c|}{Customer-to-node ratio: 67\%, PV penetration: 75\%, EV penetration 63\%}\\
\bottomrule
\end{tabular}
}
%\caption*{Q: Quadratic, LIN: Linear}
\label{TLast}
\end{table}

In examining the coordination of a very large customer base, the same feeder is examined under a significantly higher customer-to-node ratio and penetration of FRs. Specifically, the customer base was expanded, composed of 120 customers, 90 PVs, 75 EVs. This is a system of unrealistically high loading conditions, subject to massive stress. For this case, based on PF solutions, we initialize the problem and remove beforehand potentially inactive slack variables. We execute 25 simulations per inverter type; results are provided in Table \ref{TLast}.

The solution is calculated faster, owing to better tailored initialization and elimination of variables. Concerning the purely technical aspects, the  1$\Phi$ case has higher costs, owing to to the extensive re-shaping of FL/EV profiles, as well as engaging in extensive PV curtailment and reactive power injection. The 3$\Phi$ case makes extensive use of the EV/PV inverters (almost all are utilized), avoiding the need to resort to flexibility procurement almost entirely. As such, the feeder is managed much more cheaply and efficiently. In both cases, however, the DSO has access to a very large pool of residential flexibility; the subsequent technical issues are minor and the feeder is (nearly) always operated within acceptable conditions.

\subsubsection{Further expansions and limitations}

\begin{table}[b!]
\large
 \centering
\captionof{table}{Solution times (s) for different problem sizes}
\scalebox{0.8}{
\begin{tabular}{|c|c|c|c|c|}
\midrule
&  \multicolumn{4}{|c|}{Customer-to-node ratio (\%)}\\
\hline
Nodes & 10 & 20 & 30 & 40\\ 
\midrule
200 & 38.7 & 58.2 & 105.5 & 276.1\\
\hline
400 & 191.6 & 318.2 & 491.5 & 802.5\\
\hline
600 & 904.2 & 1373.4 & 2016.5 & Int\\
\hline
800 & 1617.5 & 2905.6 & 4389.9 & Int\\
\hline
\multicolumn{5}{|c|}{PV penetration: 50\%, EV penetration: 25\%}\\
\bottomrule
\end{tabular}
}
\caption*{Int: Solution manually interrupted after 10,000 seconds}
\label{TGrowing}
\end{table}

The problem size expansion can follow two directions: system expansion (nodes) and customer base expansion (controllable FRs). As such, we must explore the framework's performance with respect to two kinds of expansion. Four additional feeders (originals available in \cite{DSF_AVRA}) of 200, 400, 600 and 800 nodes, respectively, were examined (using the 3$\Phi$ inverter configuration for PVs/EVs), hosting different sizes of customer bases. A single simulation is performed per case. The customer distribution (node/phase) is random. PF-based initialization was again employed. Due to technical issues encountered (solver crashing without clear reason), the commercial solver KNITRO \cite{KNITRO} is employed instead of IPOPT. The results are presented in Table \ref{TGrowing}.

As is obvious, some of the examined setups results in very highly loaded systems, requiring the ``activation" of high percentages of FRs, thus stressing the solution process itself. As expected, increasing the system size (nodes) subsequently increases the solution time. However, if the number of controllable elements remains low (lower system stress, less variables), a locally optimal solution is (generally) achieved within acceptable time-frames for day-ahead settings. For very high penetrations of controllable elements the system is much more stressed, increasing the solution time. In fact, for larger systems hosting unrealistically high numbers of FRs, no solution is returned even after 10,000 seconds.  However, it appears that the negative impact on the solution time stems on a larger part from the number of controllable elements, rather than from the size of the system (very large system with no controllable elements are simulated very fast). For most LV distribution systems (10-30\% customer-to-node ratio), a solution can be calculated within reasonable time-frames. 

On a final note, the authors wish to re-stress that despite the several pros of the framework in and of itself, no claim is made on the quality of the chosen modelling/simulation tools. However, the exact actions to be taken for fine-tuning the solution process are out of paper scope.

\subsection{Additional cases of interest}

The results so far demonstrated the capabilities of the framework for the most common applications of optimization in LV distribution systems. However, for the sake of comprehensiveness, we re-turn our focus to the 18-node system and examine the impact (on the solution) of five more intricate modelling choices:

\subsubsection{Specialized load models}

\begin{figure}[b!] 
	\centering
	%\vspace{-0.35cm}
	\scalebox{0.16}{\includegraphics[]{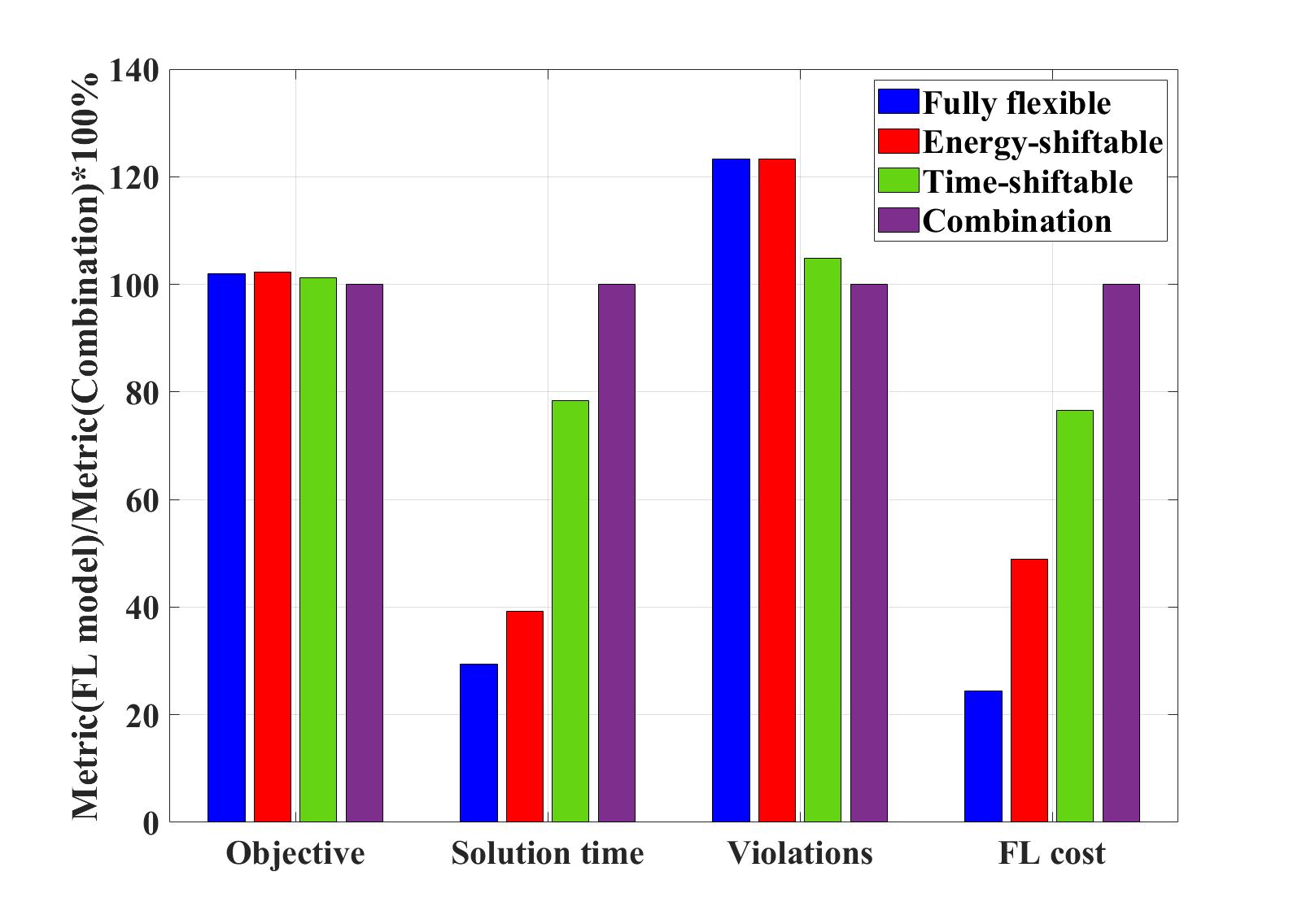}}
	\caption{Impact analysis of different FL models}
	\label{FLmodels}
\end{figure}

While the traditional flexible load model \eqref{CF} is the one most commonly used, it is still useful to perform an impact analysis for the expanded (generic) FL model. For that purpose, we consider FLs as the only controllable elements. We examine three different FL models, where 50\% of the load is fixed and the remainder 50\% is either fully flexible, energy-shiftable or time-shiftable. A fourth model (most generic) is also examined, where 25\% of the load is fixed and the remainder 75\% is equally distributed to the first three models. Results are presented in Fig. \ref{FLmodels}.

Fig. \ref{FLmodels} illustrates through various metrics the percentage difference between using the most generic FL model and different versions of it. Obviously, more refined models translate to higher FL costs, albeit to somewhat lower objective function values and total technical violations. Specifically, the first three load models result in 20\%, 19\% and 5\% more violations, respectively, than their combination, while the total FL cost is 76\%, 47\% and 22\% lower, respectively. However, the objective function values are practically the same for all FL models. Whatever small improvement is achieved comes at a cost of drastically higher solution times, with the first three models reaching an optimal solution 72\%, 61\% and 23\% faster, respectively. This is an important reason why the first FL model is the one most commonly employed. Reducing the solution time would require specialized approximation (a necessity for much larger networks), which are out of paper scope. The reader is referred to \cite{SL_Approximations} for further details.

\begin{figure}[b!] 
	\centering
	%\vspace{-0.35cm}
	\scalebox{0.57}{\includegraphics[]{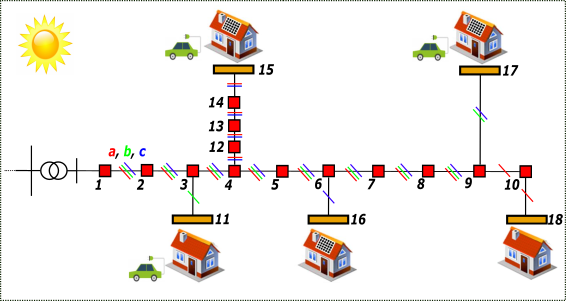}}
	\caption{Modified, multi-phase version of CIGRE LV feeder}
	\label{CigreLVMod}
\end{figure}

\subsubsection{Modelling of multi-phase lines}
While the presented results focused on the most commonly employed network models, i.e., either all-1$\Phi$ or all-3$\Phi$, the proposed framework is \textit{fully compatible with multi-phase networks} as well. An example of the framework's compatibility is the solution of the MP-OPF problem on the CIGRE LV feeder, modified to an arbitrary multi-phase system, see Fig. \ref{CigreLVMod}. As can be seen, while the main body of the feeder is 3$\Phi$, the lines directly leading to buildings are not. Specifically, the more active buildings (hosting FLs, EV, PV) are connected to the main feeder through 2$\Phi$ lines, while 1$\Phi$ lines are used for less active buildings. All devices are accordingly modified so that their phases correspond to these of their respective lines.  Modelling a multi-phase line is simply achieved by disregarding some of the parameters $R_{l,f\theta}, X_{l,f\theta}$, i.e., omitting them from the formulation after network topology processing. The phase allocation per line is completely random.

The random allocation of phases and solution of the MP-OPF was performed a number of times to validate the framework's capabilities. The main observation was that the MP-OPF always reached an optimal solution without issue. It is however worth stating that, on average, the solution time was slightly higher than that for pure 3$\Phi$ networks. This was expected, since when the number of phases is reduced the power is not distributed as well, resulting in higher thermal stressing for some lines. As previously discussed, when the system is more stressed, i.e., requires the ``activation" of additional slack variables, the solution time is generally higher. Nonetheless, this is a natural outcome of the problem setup and is not attributed to some weakness in the framework.

\subsubsection{Validation of proposed stand-alone battery model}
\begin{figure}[t!] 
	\centering
	%\vspace{-0.35cm}
	\scalebox{0.66}{\includegraphics[]{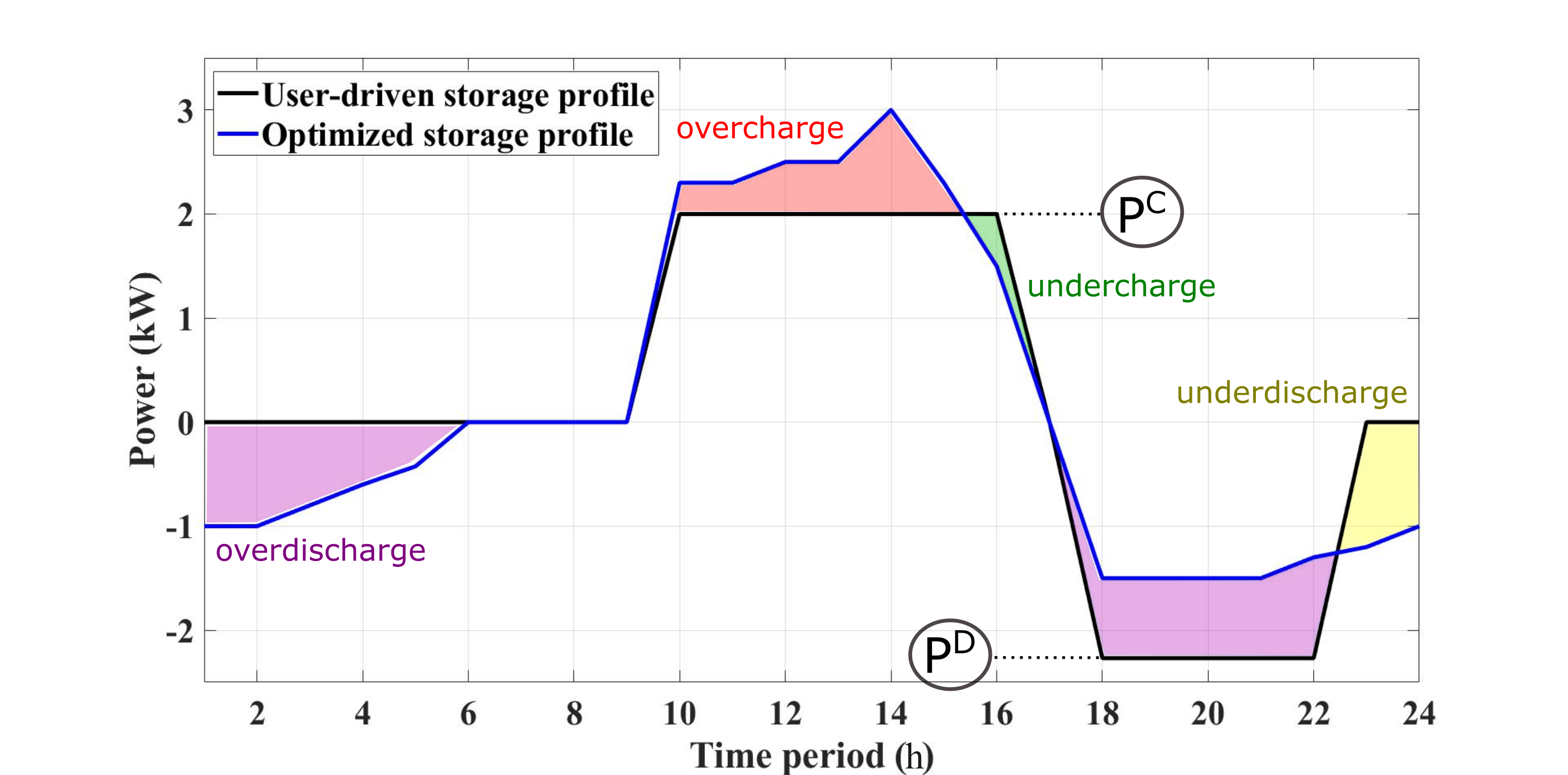}}
	\caption{Behavior of proposed battery model ($n^{C} = 0.9$)}
	\label{BSSval}
\end{figure}

The proposed stand-alone battery model is validated on the 18-node feeder. The battery, originally based on the proposed EV model, has a standard user-driven profile, sometimes charging ($P^{C}$) and sometimes discharging ($P^{D}$). Contrary to common battery models, e.g., the one presented in \cite{DOPF}, instead of ``penalizing" all charging and discharging activities, only the deviations from the  user-driven profile are remunerated. As can be seen in Fig. \ref{BSSval}, which presents the behavior of a battery at node 17, the battery originally charges at noon and discharges during the night, heuristically designed to do so by the user. Post-optimization, the battery charges more heavily at noon to counter the high PV production (avoid overvoltages), discharges more heavily at night to serve the high EV demand (avoid undervoltages). The extra stored energy is distributed to early morning hours. The simulation time remains small (comparable to what would be expected from a traditional battery model), and the battery behaves within the lines of ``normalcy". Even after optimization, the battery exchanges the exact same energy amount as originally designed, a strong point of the proposed model.  

\subsubsection{Wye-connections vs delta-connections}
\begin{figure}[t!] 
	\centering
	%\vspace{-0.35cm}
	\scalebox{0.065}{\includegraphics[]{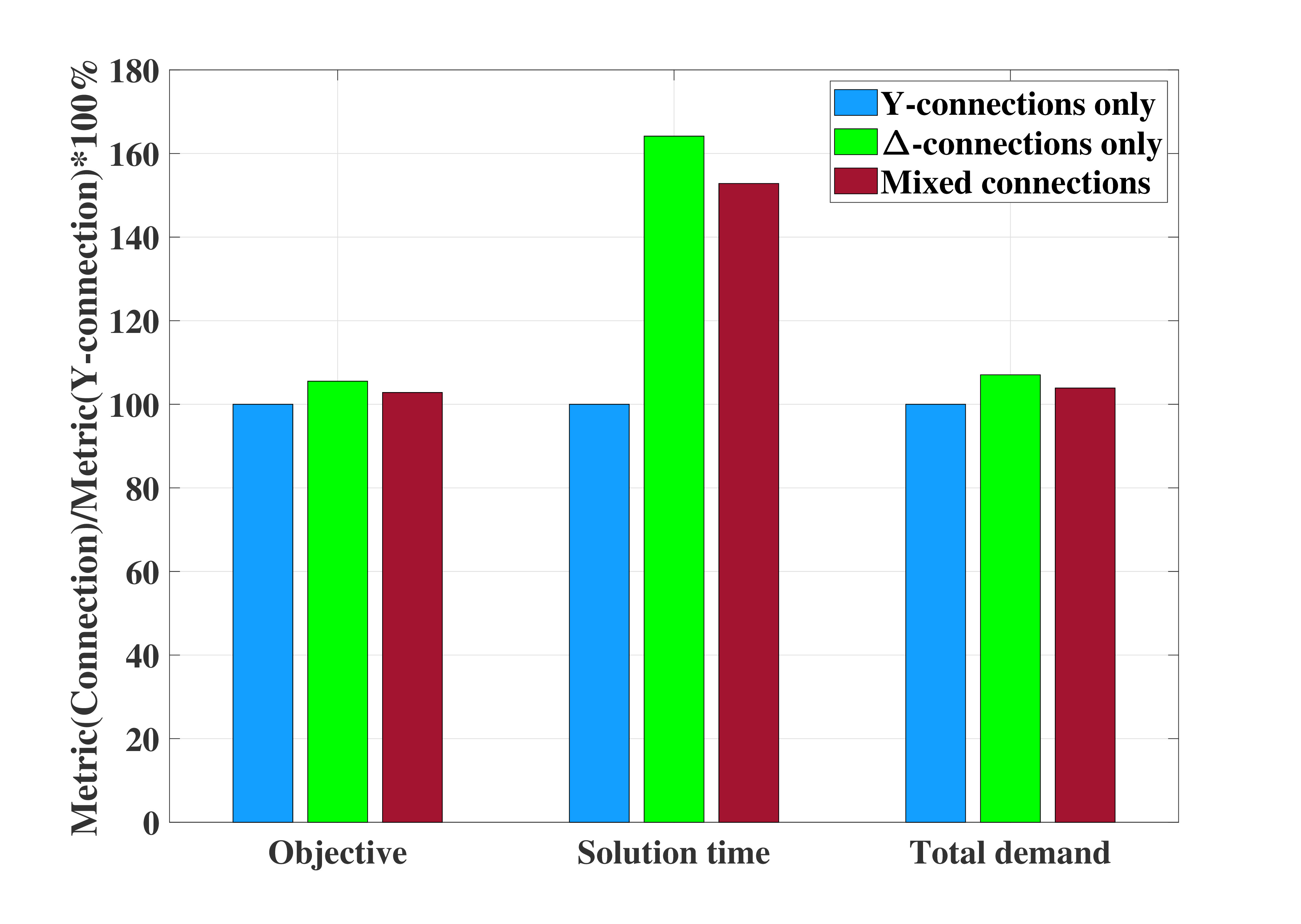}}
	\caption{Impact analysis of different connection types}
	\label{Connections}
\end{figure}

At this point it is still useful to examine how the solution is affected by the device connection type (remember that the framework is fully compatible with both connection types). For that reason, we examine three cases: the standard case, where only Y connections are assumed, the uncommon case, where only delta connections are assumed, and an intermediate case, where both connection types are assumed. All three cases are simulated through the proposed framework, and results are presented in Fig. \ref{Connections}. Delta-connected devices generally lead to higher currents (total demand), resulting in slightly higher objective function values. Systems with more delta connections may also require more time to solve, since not only do the constraints become more intertwined, but the increased demand also leads to higher system stress, which, as has been shown, may slow down the solution time. However, such cases are rare in practice, as most residential loads are in fact wye-connected.

\subsubsection{Evaluation of the V2G capability}
Though not as common, some EVs also have the V2G capability. This capability has limited impact during weekdays: since the EV is not present for a big part of the day (owner is away), this cripples the battery's potential for providing flexibility, as there are no clear opportunities for it to charge and serve any possible need for self-consumption during nighttime. On the other hand, the consumption re-distribution capability has limited potential during weekends, where the limited need for any charging severely constrains the potential of 3$\Phi$ inverters. The aforementioned observations are confirmed by simulations with both inverter configurations. The best case scenario would be to consider an EV with a 3$\Phi$ inverter configuration and the V2G capability, thus this setup is rare itself.

\begin{figure}[t!] 
	\centering
	%\vspace{-0.35cm}
	\scalebox{0.14}{\includegraphics[]{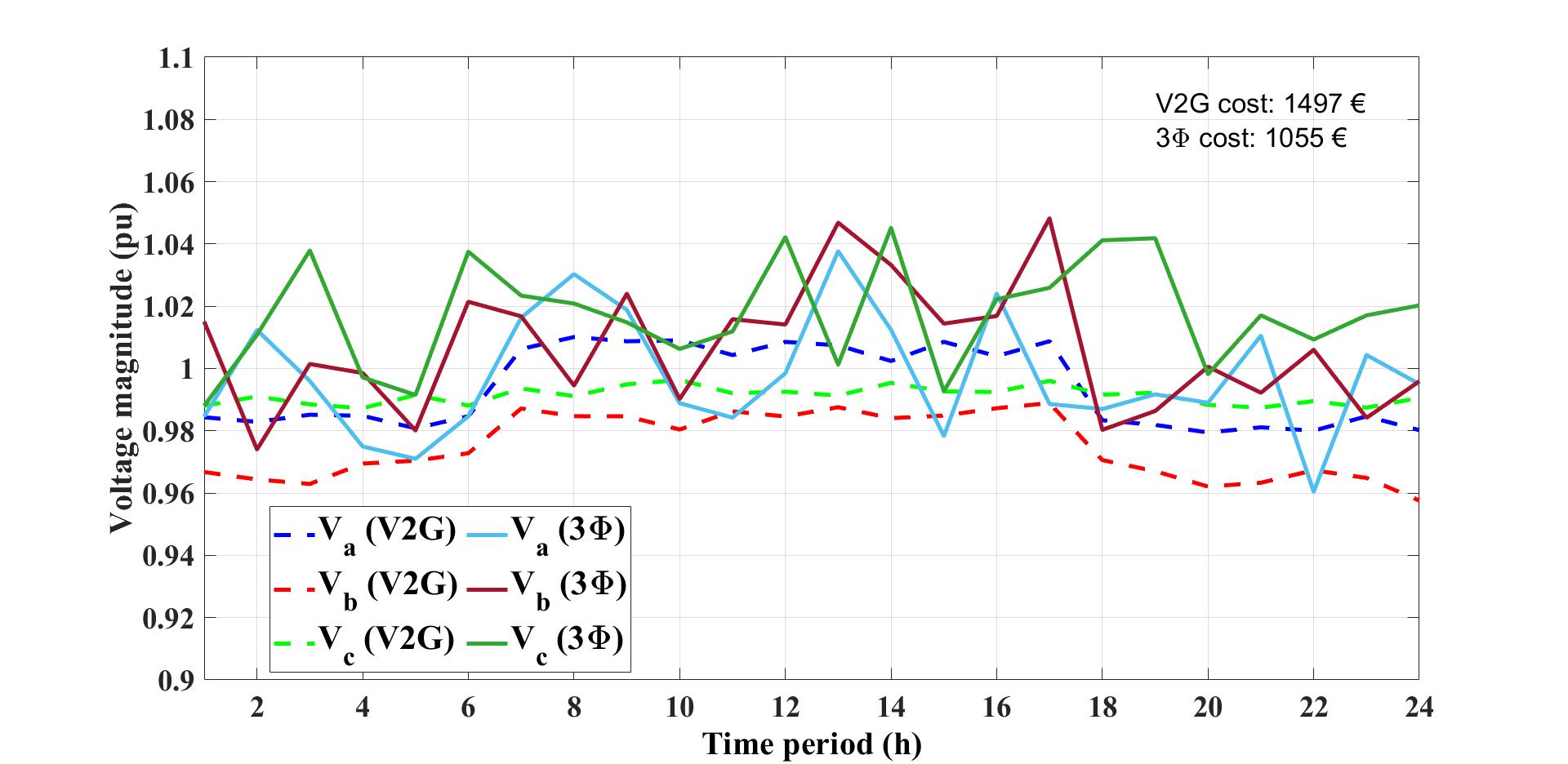}}
	\caption{18-node feeder, average voltage magnitude per phase: V2G vs power re-distribution comparison}
	\label{V2Gval}
\end{figure}

To make a fair comparison, we examine an idealized setup where the EV must charge its usual amount, but without temporal restrictions, i.e., the EV may interact with the grid at all times. We compare the performance of the V2G capability (1$\Phi$ configuration) against the consumption re-distribution capability (3$\Phi$ configuration). We also set the exporting price to 20 \euro/kW to motivate system self-sufficiency (see also Case 3). As can be seen in Fig. \ref{V2Gval}, the 3$\Phi$ inverter configuration beats the V2G capability by a clear margin. Under the former, the average per-phase voltage level is generally more elevated and fluctuates more closely to unity. In addition, the remuneration cost with the 3$\Phi$ inverter configuration is significantly lower. When the V2G capability is employed, the phase to which the EV is connected must engage in drastic action (charging alteration and discharging), utilizing two different flexibility services (thus driving up the cost) and only partially managing the problem. With the 3$\Phi$ inverter configuration, the EV must only alter its charging profile. By constantly balancing its operation between the three phases, the flexibility cost is kept lower. No single phase disproportionately affects the others, as the 3$\Phi$ inverter configuration ensures that the interactions between phases is more balanced and more harmonious.

\section{Conclusions}

The authors have constructed a versatile MP-OPF framework that serves two key functions: it proposes and compares the performance of state-of-the-art device models for unlocking the flexibility potential of smart distribution grids and it provides up-to-date guidelines with respect to how each of the most commonly employed (or ignored) modelling choices (concerning the loads, the network and the degrees of controlability) affect the quality and reliability of the solution. The reported results can guide researchers into picking proper equipment models depending on their respective needs. The formulation also scales well for larger systems (under proper conditions). As such, it can serve as a solid basis for approaches aiming specifically at scalability; its results can also be safely contemplated by the DSO for hours-ahead use.

The impact of different versions of common FR devices was analyzed, based on novel and realistic models. Namely, the authors proposed flexible ZIP load models (profile alteration affects Z, I, P components), both $1 \Phi$ and $3 \Phi$ (balancing) versions of PVs with realistic associated costs, reactive capabilities and PCC curves and both $1 \Phi$ and $3 \Phi$ (balancing) versions of EVs with the added novelty of building on the original customer-desired profile, rather than determining an original profile altogether. The proposed models can be used in devising new approaches for unlocking the full potential of FRs to manage violated constraints in distribution systems. 

In the future, the authors plan to extend their MP-OPF framework to to address the issues of uncertainty, an area were, due to the huge computational challenge, the research is scarce and often limited to small systems.

\bibliographystyle{unsrt}
%\bibliography{ref}

\begin{thebibliography}{10}

\bibitem{SGVision}
{H. Farhangi}.
\newblock {``The Path of the Smart Grid"}.
\newblock {\em IEEE Power and Energy Mag.}, 8(1):18--20, 2010.

\bibitem{Passive_to_Active}
{A. Keane, L. Ochoa, C. Borges, et al}.
\newblock {``State-of-the-Art Techniques and Challenges Ahead for Distributed
  Generation Planning and Optimization"}.
\newblock {\em IEEE Trans. Power Syst.}, 28(2):1493--1502, 2013.

\bibitem{DSF_AVRA}
{I.I. Avramidis, V.A. Evangelopoulos, P.S. Georgilakis, and N. Hatziargyriou}.
\newblock {``Demand side flexibility schemes for facilitating the high
  penetration of residential distributed energy resources"}.
\newblock {\em IET Gener. Transm. Distr.}, 12(18):4079--4088, 2018.

%I. Kockar, and G. Ault
\bibitem{DOPF}
{S. Gill, et al}.
\newblock {``Dynamic Optimal Power Flow for Active Distribution Networks"}.
\newblock {\em IEEE Trans. Power Syst.}, 29(1):121--131, 2014.

\bibitem{4-wire_model}
{M.J.E. Alam, K. M. Muttaqi, and D. Sutanto}.
\newblock {``A Three-Phase Power Flow Approach for Integrated 3-Wire MV and
  4-Wire Multigrounded LV Networks With Rooftop Solar PV"}.
\newblock {\em IEEE Trans. Power Systems}, 28(2):1728--1737, 2013.

\bibitem{Efficient_OLTC}
{S.R. Shukla, S. Paudyal, and M.R. Almassalkhi}.
\newblock {``Efficient Distribution System Optimal Power Flow With Discrete
  Control of Load Tap Changers"}.
\newblock {\em IEEE Trans. Power Systems}, 34(4):2970--2979, 2019.

\bibitem{Reconfiguration_FC}
{F. Capitanescu, L.F. Ochoa, H. Margossian, and N.D. Hatziargyriou}.
\newblock {``Assessing the Potential of Network Reconfiguration to Improve
  Distributed Generation Hosting Capacity in Active Distribution Systems"}.
\newblock {\em IEEE Trans. Power Systems}, 30(1):346--356, 2015.

\bibitem{OPF_SmartDistributionFeeders}
{S. Paudyal, C. Ca\~{n}izares, and K. Bhattacharya}.
\newblock {``Optimal Operation of Distribution Feeders in Smart Grids"}.
\newblock {\em IEEE Trans. Industrial Electronics}, 58(10):4495 -- 4503, 2011.

%P. Aristidou, and Gabriela Hug
\bibitem{Karagiannopoulos_PSCC}
{S. Karagiannopoulos, et al}.
\newblock {``A Centralised Control Method for Tackling Unbalances in Active
  Distribution Grids"}.
\newblock In {\em PSCC}, June, 2018.

\bibitem{CVR_Unbalanced}
{L. Gutierrez-Lagos and L.F. Ochoa}.
\newblock {``OPF-based CVR Operation in PV-Rich MV-LV Distribution Networks"}.
\newblock {\em IEEE Trans. Power Systems}, 34(4):2778 -- 2789, 2019.

\bibitem{MPOPF_CurrentImbalance}
{A. O'Connell and Andrew Keane}.
\newblock {``Multi-Period Three-Phase Unbalanced Optimal Power Flow"}.
\newblock In {\em ISGT Europe}, October, 2014.

\bibitem{Local_improvement_EV_PV}
{F. Marra, G.Y. Yang, Y.T. Fawzy, et al}.
\newblock {``Improvement of Local Voltage in Feeders With Photovoltaic Using
  Electric Vehicles"}.
\newblock {\em IEEE Trans. Power Systems}, 28(3):3515--3516, 2013.

\bibitem{Rolling_MPOPF}
{A. O'Connell, D. Flynn, and A. Keane}.
\newblock {``Rolling Multi-Period Optimization to Control Electric Vehicle
  Charging in Distribution Networks"}.
\newblock {\em IEEE Trans. Power Systems}, 29(1):340--348, 2014.

\bibitem{Central_and_Local_DS}
{S. Weckx, C. Gonzalez, and J. Driesen}.
\newblock {``Combined Central and Local Active and Reactive Power Control of PV
  Inverters"}.
\newblock {\em IEEE Trans. Sustainable Energy}, 5(3):776--784, 2014.

\bibitem{PV_4wire}
{X. Su, M.A.S. Masoum, and P.J. Wolfs}.
\newblock {``Optimal PV Inverter Reactive Power Control and Real Power
  Curtailment to Improve Performance of Unbalanced Four-Wire LV Distribution
  Networks"}.
\newblock {\em IEEE Trans. Sustainable Energy}, 5(3):967--977, 2014.

\bibitem{PhaseBalancing}
{S. Weckx and J. Driesen}.
\newblock {``Load Balancing with EV Chargers and PV Inverters in Unbalanced
  Distribution Grids"}.
\newblock {\em IEEE Trans. Sustainable Energy}, 6(2):635 -- 643, 2015.

\bibitem{OptimalBSSDispatchUnbalanced}
{J.D. Watson, N.R. Watson, and I. Lestas}.
\newblock {``Optimized Dispatch of Energy Storage Systems in Unbalanced
  Distribution Networks"}.
\newblock {\em IEEE Trans. Sustainable Energy}, 9(2):639 -- 650, 2017.

\bibitem{BSS_LV}
{P. Fortenbacher, J.L. Mathieu, and G. Andersson}.
\newblock {``Modeling and Optimal Operation of Distributed Battery Storage in
  Low Voltage Grids"}.
\newblock {\em IEEE Trans. Power Systems}, 32(6):4340--4350, 2017.

\bibitem{CVR_Unbalanced_High_PV}
{Q. Zhang, K. Dehganpour, and Z. Wang}.
\newblock {``Distributed CVR in Unbalanced Distribution Systems with PV
  Penetration"}.
\newblock {\em IEEE Trans. Smart Grid}, 10(5):5308--5319, 2019.

\bibitem{Model-Free_Unbalanced}
{M.D. Sankur, R. Dobbe, A. von Meier, and D.B. Arnold}.
\newblock {``Model-Free Optimal Voltage Phasor Regulation in Unbalanced
  Distribution Systems"}.
\newblock {\em IEEE Trans. Smart Grid}, 11(1):884--894, 2020.

\bibitem{ZIP_CVR}
{S. Hossan, H.M.M. Maruf, and B. Chowdhury}.
\newblock {``Comparison of the ZIP load model and the exponential load model
  for CVR factor evaluation"}.
\newblock In {\em IEEE Power \& Energy Society General Meeting}, July, 2017.

\bibitem{Impact_of_load_model_unbalanced}
{B. Bletterie, A. Latif, et al}.
\newblock {``On the impact of load modelling on distribution network studies"}.
\newblock In {\em ISGT Europe}, September, 2017.

\bibitem{OPF_Unbalanced_Low}
{L. Gan, N. Li, U. Topcu, and S. Low}.
\newblock {``Optimal Power Flow in Distribution Networks"}.
\newblock In {\em Proceedings 52nd IEEE Conf. on Decision and Control}, 2013.

\bibitem{OPF_Unbalanced_DeltaLow}
{C. Zhao, E. Dall'Anese, and S. Low}.
\newblock {``Optimal Power Flow in Multiphase Radial Networks with Delta
  Connections"}.
\newblock In {\em IREP Bulk Power Systems Dynamics and Control Symposium},
  August, 2017.

\bibitem{BranchFlowPartOne}
{M. Farivar and S. Low}.
\newblock {``Branch Flow Model: Relaxations and Convexification-Part I"}.
\newblock {\em IEEE Trans. Power Systems}, 28(3):2554 -- 2564, 2013.

%W. Wu, and B. Zhang
\bibitem{CapacityAssessmentUnbalanced}
{X. Chen, et al}.
\newblock {``Robust Capacity Assessment of Distributed Generation in Unbalanced
  Distribution Networks Incorporating ANM Techniques"}.
\newblock {\em IEEE Trans. Sustainable Energy}, 9(2):651 -- 663, 2018.

\bibitem{ReactiveDispatchDistribution}
{B.A. Robbins and A.D. Dominguez-Garcia}.
\newblock {``Optimal Reactive Power Dispatch for Voltage Regulation in
  Unbalanced Distribution Systems"}.
\newblock {\em IEEE Trans. Power Systems}, 31(4):2903 -- 2913, 2016.

\bibitem{ACOPF_DS}
{K. Christakou, D.C. Tomozei, J.Y. Le Boudec, and M. Paolone}.
\newblock {``AC OPF in radial distribution networks – Part I: On the limits
  of the branch flow convexification and the alternating direction method of
  multipliers"}.
\newblock {\em Electric Power Systems Research}, 143:438--450, 2017.

\bibitem{UnbalancedWithNeutralSDP}
{M. Usman, A. Cervi, M. Coppo, et al}.
\newblock {``Centralized OPF in Unbalanced Multi-Phase Neutral Equipped
  Distribution Networks Hosting ZIP Loads"}.
\newblock {\em IEEE Access}, 7:177890 -- 177908, 2019.

\bibitem{UOPF_MILP_MIQCP}
{L. Gutierez-Lagos, M.Z. Liu, and L.F. Ochoa}.
\newblock {``Implementable Three-Phase OPF Formulations for MV-LV Distribution
  Networks: MILP and MIQCP"}.
\newblock In {\em ISGT Latin America}, September, 2019.

\bibitem{OptimalRestorationUnbalanced}
{J.C. Lopez, J.F. Franco, J. Rider, and R. Romero}.
\newblock {``Optimal Restoration/Maintenance Switching Sequence of Unbalanced
  three-Phase Distribution Systems"}.
\newblock {\em IEEE Trans. Smart Grid}, 9(6):6058 -- 6068, 2018.

\bibitem{EDN_Book}
Thomas~Allen Short.
\newblock {\em Electric Power Distribution Handbook, 2nd Edition}.
\newblock CRC Press, 2014.

\bibitem{SL_Approximations}
{I.I. Avramidis, F. Capitanescu, and G. Deconinck}.
\newblock {``Practical Approximations and Heuristic Approaches for Managing Shiftable Loads in the Multi-Period Optimal Power Flow Framework"}.
\newblock {\em Electric Power Systems Research}, 190: 106864, 2021.

\bibitem{EV_4Q}
{J. Wang, E.Y. Ucer, S. Paudyal, et al}.
\newblock {``Distribution Grid Voltage Support with Four Quadrant Control of
  Electric Vehicle Chargers"}.
\newblock In {\em IEEE PES General Meeting}, August, 2020.

\bibitem{PF_5phase_Nando}
{R.M Ciric, A. Padilha, and L.F. Ochoa}.
\newblock {``Power Flow in Four-Wire Distribution Networks-General Approach"}.
\newblock {\em IEEE Trans. Power Systems}, 18(4):1283 -- 1290, 2003.

\bibitem{CigreLV}
{ K. Strunz, E. Abbasi, R. Fletcher, et al}.
\newblock {``Benchmark Systems for Network Integration of Renewable and
  Distributed Energy Resources"}.
\newblock {\em CIGRE Task Force C6.04}, 4:4--6, 2014.

\bibitem{IPOPT}
A.~W{\"a}chter and L.~T. Biegler.
\newblock {``On the Implementation of a Primal-Dual Interior Point Filter Line
  Search Algorithm for Large-Scale Nonlinear Programming"}.
\newblock {\em Mathematical Programming}, 106(1):25--57, 2006.

\bibitem{GAMS}
{B.A. McCarl, GAMS user guide, version 23.8, 2012}, {www.gams.com}.

\bibitem{KNITRO}
{R.H. Byrd, J. Nocedal, and R.A. Waltz}.
\newblock {``KNITRO: An integrated package for nonlinear optimization"}.
\newblock {\em Large-Scale Nonlinear Optimization}, vol. 83, Springer:35--59,
  2006.
 

\end{thebibliography}

\end{document}